\documentclass[fp,twocolumn]{jpsj3}
\usepackage{txfonts}
\usepackage{filecontents}
\usepackage{xcolor}
\usepackage{mathtools}
\usepackage{physics}
\usepackage{ulem}
\usepackage{tikz}
\usepackage{cite}
\usepackage{siunitx} 
\usetikzlibrary{positioning}
\newcommand{\bes}{\begin{subequations}}
\newcommand{\ees}{\end{subequations}}
\newcommand{\be}{\begin{equation}}
\newcommand{\ee}{\end{equation}}

\definecolor{tu_orange}{HTML}{f28400}
\definecolor{tdo_darkgreen}{HTML}{839A00}
\definecolor{tdo_green}{HTML}{83B818}

\usepackage[
    acronym,
    hyperfirst = false
]{glossaries}
\glsdisablehyper
\newacronym{cst}{CST}{continuous similarity transformation}
\newacronym{cut}{CUT}{continuous unitary transformation}
\newacronym{scnlswt}{scNLSWT}{self-consistent nonlinear spin-wave theory}
\newacronym{nlswt}{NLSWT}{nonlinear spin-wave theory}
\newacronym{swt}{SWT}{spin-wave theory}
\newacronym{sbmft}{SB-MFT}{Schwinger boson mean-field theory}
\newacronym{dmmft}{DM-MFT}{Dyson-Maleev boson mean-field theory}

\newacronym{dmrg}{DMRG}{density-matrix renormalization group}

\newacronym{qpc}{qpc}{quasi-particle-conserving}
\newacronym{0n}{0n}{ground-state-separating}

\newacronym{rod}{ROD}{residual off-diagonality}
\usepackage[capitalise]{cleveref}
\crefname{section}{Sec.}{sections}
\makeatletter
\AddToHook{cmd/appendix/before}{\def\cref@section@alias{appendix}\def\cref@subsection@alias{appendix}}
\makeatother

\crefname{appendix}{App.}{appendizes}
\crefname{appendix}{App.}{Appendizes}

\newcommand{\sumnn}{\sum_{\langle i,j\rangle}}
\newcommand{\kk}{\mathbf{k}}
\newcommand{\kkm}{\mathbf{-k}}

\newcommand{\lr}[1]{\left( #1 \right)}

\newcommand{\lrg}[1]{\left\{#1\right\}}
\newcommand{\lara}[1]{\langle #1 \rangle}

\newcommand{\nbi}[1]{\hat{#1}_i^\dagger \hat{#1}_i}
\newcommand{\nbj}[1]{\hat{#1}_j^\dagger \hat{#1}_j}

\newcommand{\nbdij}[1]{\hat{#1}_i^\dagger \hat{#1}_j^\dagger}
\newcommand{\nbij}[1]{\hat{#1}_i \hat{#1}_j}

\newcommand{\sumk}{\sum_{\mathbf{k}}}

\newcommand{\ak}{\hat a_\mathbf{k}}
\newcommand{\bk}{\hat b_\mathbf{k}}
\newcommand{\akm}{\hat a_\mathbf{-k}}
\newcommand{\bkm}{\hat b_\mathbf{-k}}

\newcommand{\gam}{\gamma_\mathbf{k}}

\newcommand{\alk}{\hat{\alpha}_\mathbf{k}}
\newcommand{\bek}{\hat{\beta}_\mathbf{k}}

\title{Control of the N\'eel vector in the quantum antiferromagnetic honeycomb  lattice}

\author{Asliddin Khudoyberdiev\thanks{asliddin.khudoyberdiev@tu-dortmund.de}, Dag-Björn Hering, Vanessa Sulaiman, and Götz S. Uhrig}
\inst{Condensed Matter Theory, TU Dortmund University, Otto-Hahn-Straße 4, 44227 Dortmund, Germany\\
} 

\abst{The switching of antiferromagnetic order and its efficient control promise to enable ultrafast manipulation of data  and large storage capacity. Recently, the time-dependent Schwinger boson mean-field theory has been successfully developed to study the N\'eel vector switching in hypercubic antiferromagnetic lattices. In the present article,  we aim at demonstrating that the approach is a well-justified framework to capture the essentials of the switching process, even in low-symmetry quantum antiferromagnets. To this end,  we show the possibility of the sublattice magnetization reorientation  in the quantum antiferromagnetic  honeycomb lattice. First, equilibrium properties of the honeycomb lattice are analyzed using the Schwinger boson mean-field theory and compared to the continuous similarity transformation method to justify the applicability of the approach. Then, the Schwinger boson mean-field theory is employed for switching process. We provide a comprehensive answer to the question what the threshold switching fields are when the coordination number of the lattice is varied. Indeed, the results of the study reveal a correspondence between lattice structures and the threshold fields by comparing them for the square and the
simple cubic lattices and the honeycomb lattice.  The findings of the present article extend the foundation for future theoretical and computational advancements in the field of antiferromagnetic switching. These advancements are of particular relevance for the development of ultrafast spintronic or magnonic devices.}

\begin{document}
\maketitle

\section{Introduction}

The ultrafast manipulation of magnetic order in spintronics has significance for the advancement of future terahertz data processing. Assuming precise control of the N\'eel vector, antiferromagnets are promising candidates in this respect because of their ultrafast spin dynamics, while their potential for the high storage density provides an additional advantage~\cite{jungw16}. Experimental observations have already demonstrated advancements in the manipulation and control of antiferromagnetic order~\cite{ wadle16,meiner18,jourd25,zhou25}, and ongoing measurements are providing further insights into these developments. For instance, very recently Behovits \textit{et al.}~\cite{Behov25} demonstrated the direction-controlled, non-thermal, robust energy-efficient and ultrafast rotation of the N\'eel vector by $\pm 90^{\circ}$ in the antiferromagnet Mn$_2$Au via terahertz electric field pulses at room temperature. While the spins are essentially quantum, it is not clear whether a quantum description is strictly necessary and capable to describe the magnetization switching and its control~\cite{bolsm23,yarmo25,zelez14}, or arguments in terms of classical spin  would be adequate as well~\cite{zelez14,gomon10,gomon16,liao20,zhang26}. We note that to the best of our knowledge, a consistent combination of experiment and quantum theory
has not been demonstrated for the magnetization switching in antiferromagnets.

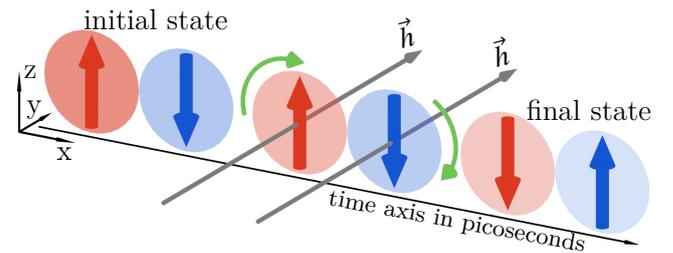
\begin{figure}
 \begin{tikzpicture}[
    every node/.style={
        font=\fontsize{11}{11}\selectfont, 
    }
]
\node at (-2.4,1.6) {initial state};
\node at (3.3,0.4) {final state};
\node at (-4.0,0.4) {y};
\node at (-3.6,-0.18) {x};
\node at (-4.05,0.9) {z};
\node [rotate=15, anchor=south west] at (2.0,0.81) {$\vec{h}$};
\node[rotate=15, anchor=south west] at (0.75,1.06) {$\vec{h}$};
\node[rotate=-10.5, anchor=south west,font=\small] at (-0.4,-1.0) {time axis in picoseconds};
\node at (0,0) {\includegraphics[width=\columnwidth]{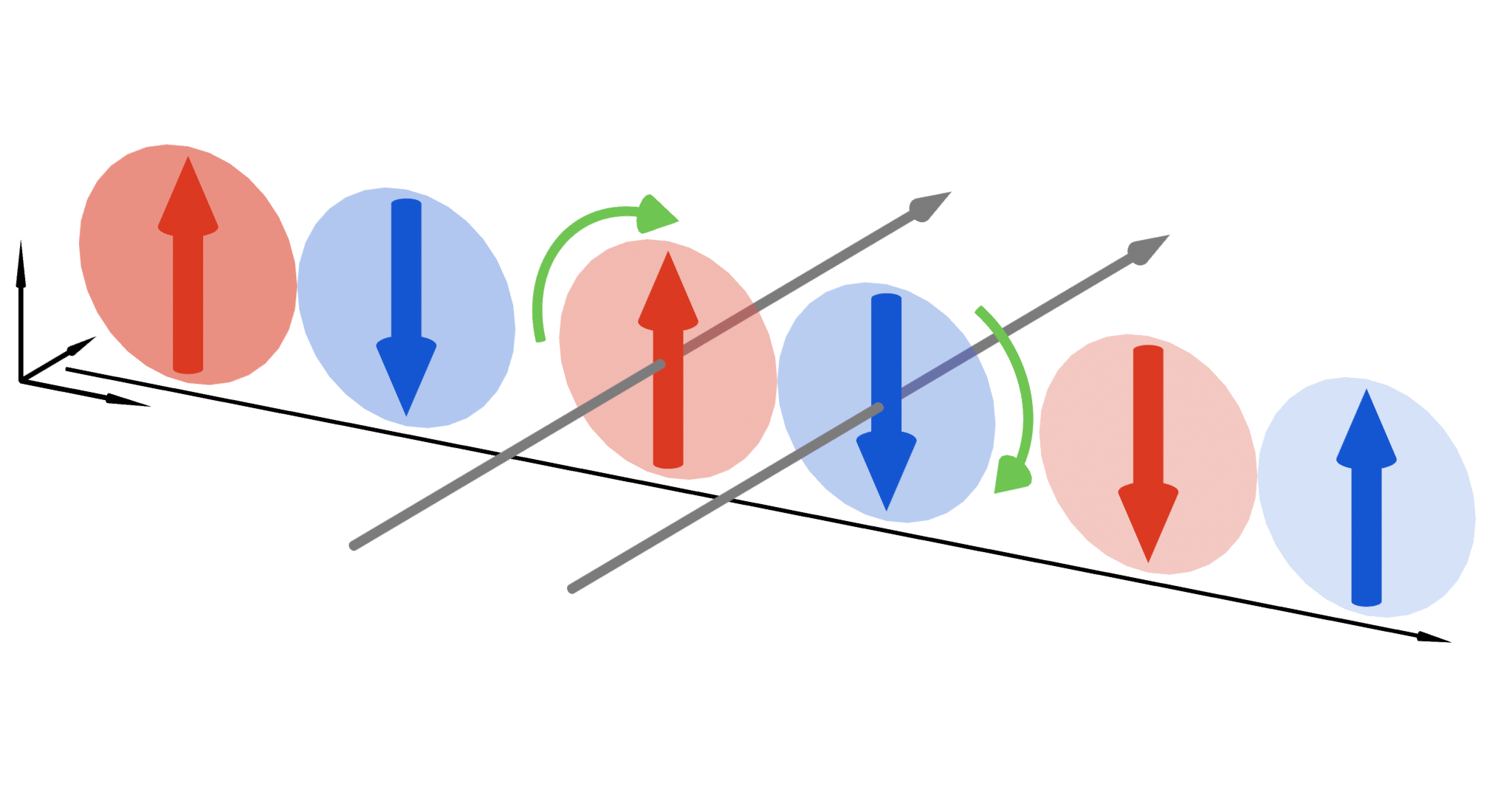}};
\end{tikzpicture}
\vspace{-1.7cm}
\caption{This classical illustration shows how the sublattice magnetization is manipulated. The initial state, with two linked red and blue sublattices, is shown on the left. The arrows inside the circles indicate the direction of the sublattice magnetization. A uniform magnetic field (gray arrows) is applied to switch the order (middle), and the final state is obtained (last two circles on the right). The round green arrows indicate the direction of the Larmor precession of the spins on the sublattice.} \label{fig:illustswitch}
\end{figure}

Recently, it was shown that the time-dependent \gls{sbmft} can describe the switching of the order in  anisotropic qunatum antiferromagnets by means of an external magnetic field~\cite{bolsm23, khudoy24}, as illustrated in \cref{fig:illustswitch}. The approach captures the dephasing effect which slowly brings the system to a steady-state after switching. The results also show that the N\'eel vector can be reversed in the ultrafast regime under low external fields through the mechanism of exchange enhancement~\cite{khudoy25}. The incorporation of the spin-lattice relaxation leads to an exponentially fast convergence to the steady-state following full ultrafast switching~\cite{khudoy25scipost}.  Furthermore, the complementary exact diagonalization methods in small clusters also supported the results of the aforementioned approach~\cite{johann26}. So far, the quantum switching process has been analyzed on hypercubic lattices in 2d and 3d. The reorientation of the sublattice magnetisation using the \gls{sbmft} in more complex lattices with low symmetries is a subject requiring further study.

In this paper, we report the analysis of the  switching in the spin-1/2 two-dimensional quantum antiferromagnetic honeycomb lattice, with a focus on significant quantum effects. The purpose of this study is to demonstrate the capability of the Schwinger boson representation to capture the switching process in a hexagonal lattice by encompassing the full spectrum of anisotropy in the XXZ Heisenberg model. 
To this end the equilibrium state without an external magnetic field is analyzed first.
In this context, we take the opportunity to also present results obtained using \gls{cst}, which are founded on a previous successful application in the case of the XXZ model on the square lattice~\cite{Walther2023}. 
However, it is important to emphasize that the Schwinger boson approach and the \gls{cst} based on the Dyson-Maleev representation differ conceptually. 
Only the Schwinger boson approach is used here to describe switching processes, and
the integration of the \gls{cst} to switching is left to future research.
We provide general formulas for the minimal strength of the field required to switch the orientation of the spins depending on the anisotropy in any bipartite quantum antiferromagnet. 
The simulations will be conducted in the ground state of a large, but finite system.
Finite temperature analyses are left for subsequent studies.

So far, the \gls{sbmft} was successful to capture the novel phases in the frustrated quantum $J_1$-$J_2$ Heisenberg antiferromagnet on
the honeycomb lattice~\cite{zhang13}. The analysis of the model showed an increased N\'eel stability relative to the classical case~\cite{mattss94}. Furthermore, the approach was employed to study the $J_1$-$J_2$-$J_3$ Heisenberg model on the same lattice~\cite{carba11},  showing a quantum spin liquid phase for $S=1/2$ (which disappears for $S=1$) stabilized between the N\'eel, spiral, and collinear antiferromagnetic phases~\cite{meri18}.  The theory was also applied to investigate the biquadratic bilinear Heisenberg antiferromagnetic model in a honeycomb lattice at both zero and low temperatures, thereby depicting the ordered and disordered states, respectively~\cite{moura13}.  Based on these  studies, we believe that the Schwinger boson approach is a well-suited theory for considering the antiferromagnetic honeycomb lattice. Due to the small coordination number of the honeycomb lattice, the quantum fluctuations are stronger than those in the square
and simple cubic lattices. Consequently, we expect a quantitative change in the magnitude of the required switching fields.

The structure of the paper is as follows. The model and the details of the \gls{sbmft} are provided in the next Section together with a brief introduction to \gls{cst}. 
The equilibrium properties of the system are analyzed in \cref{sec:equil}
and to corroborate the \gls{sbmft} results also results of \gls{cst} are shown. In \cref{sec:switch}, we outline the derivation of the non-equilibrium equations. 
\cref{sec:results} is devoted to results and discussion of the switching process. Finally, we draw our conclusions in  \cref{sec:con}.

\section{Model and Method}
Our approach is based on the anisotropic quantum antiferromagnetic Heisenberg model on the  honeycomb lattice with nearest neighbor interactions between the two magnetic sublattices. The Hamiltonian of the system reads ($\hbar=1$)
\begin{equation} \label{eq:Heisenberg}
    \mathcal{\hat{H}}_\text{0}= J \sum_{\langle i,j\rangle}\left[\frac{\chi}{2}(\hat{S}_i^x\hat{S}_j^x+\hat{S}_i^y\hat{S}_j^y)+\hat{S}_i^z\hat{S}_j^z\right] ,
\end{equation}
where $J$ is the exchange coupling constant, and $\chi\in[0,1]$ the anisotropy parameter. Throughout the paper, we denote operators by ``hats'' to facilitate the distinction between physical operators and scalar variables for the reader.

\subsection{Schwinger boson representation}

\begin{figure}
\includegraphics[width=0.98\columnwidth]{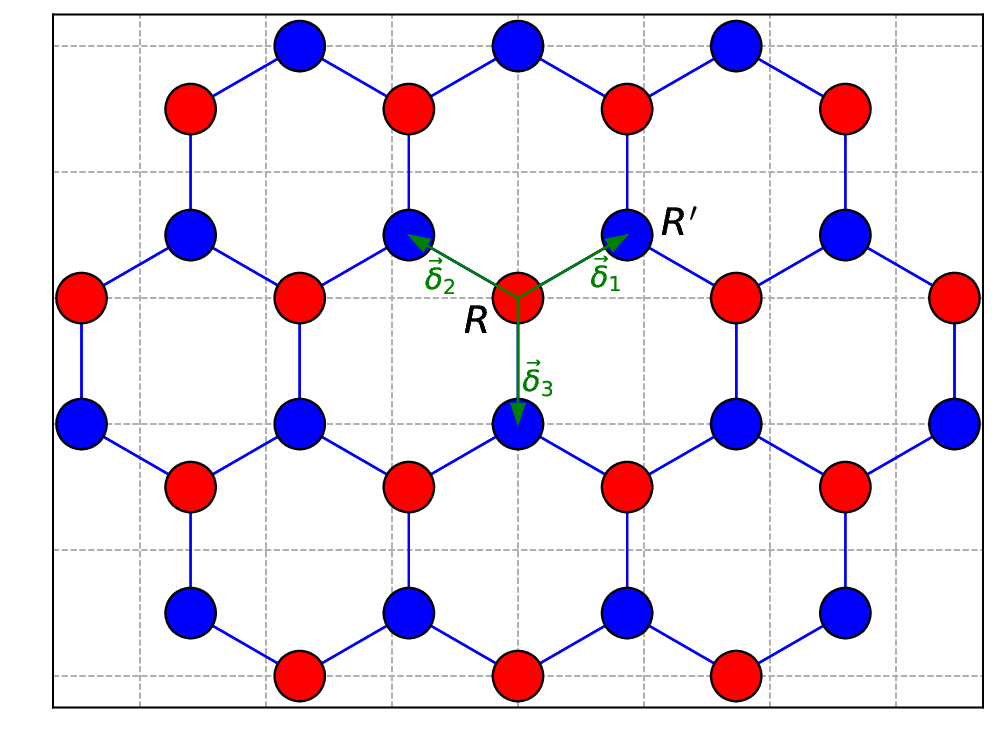}
\caption{Honeycomb lattice with two nonequivalent sites.    
The sublattice $R$ correspond to site $i$ and sublattice $R'$ correspond to site $j$. The vectors $\boldsymbol{\delta}_l$ ($l=1,2,3$) connect the  site $R$ to its three neighbouring $R'$ sites.} 
\label{fig:illus}
\end{figure}

The Schwinger boson representation does not require specifying a preferential
direction and hence has the great merit to allow  for arbitrary orientations of the sublattice magnetization~\cite{arova88,auerb94,pires21, zhang22}. The representation has been successfully employed for the equilibrium  dynamics of quantum spin models~\cite{auerb88,ng11,schuck18}
and for switching processes in antiferromagnets~\cite{bolsm23,khudoy24,khudoy25}.  Therefore, we continue to use the SU(2) 
Schwinger boson representation in the present work. 
The representation lies on replacing the spin operators through the bosonic ladder operators that 
reproduce the behavior of the spin as 
\be
\hat{\mathbf{S}}_i=\frac{1}{2}\begin{pmatrix}
\hat{a}^\dagger_i  \hat{b}^\dagger_i
\end{pmatrix}\boldsymbol{\sigma}
\begin{pmatrix}
  \hat{a}_i \\
  \hat{b}_i \\
\end{pmatrix}, 
\ee
where $\boldsymbol{\sigma}$ is the Pauli vector. However, the bosonic Hilbert space is infinitely large and thus too large to describe
the (2$S$+1)-dimensional spin Hilbert space given the spin quantum number $S$.
In order to remain in the physical sector after enlarging the Hilbert space, one has to restrict
the bosonic Hilbert space with the constraint on the number of Schwinger bosons at each site
\be \label{eq:constr}
\nbi{a}+\nbi{b}=2S, \qquad \nbj{a}+\nbj{b}=2S,
\ee
where $i\in R$, $j\in R'$ (\cref{fig:illus}). The sublattice magnetization follows from the mean-occupation number of bosons as
\begin{equation}
\label{eq:magnetization}
   m_R = \frac{1}{2}\lr{\langle \nbi{a} \rangle  - \langle \nbi{b}\rangle} \quad \text{or} \quad m_{R'} = \frac{1}{2}\lr{\langle \nbj{a} \rangle  - \langle \nbj{b}\rangle},
\end{equation}
where $\langle \mathcal{\hat O}\rangle=\Tr(\hat{\rho}\hat{\mathcal{O}})$ means the expectation value of an operator $\mathcal{\hat{O}}$ for a given density matrix $\hat{\rho}$. Throughout the paper, we use ``sublattice magnetization'' and ``magnetization'' interchangeably. However, both phrases always refer to the magnetization $m:=m_R$ (the 
$z$ axis is chosen as the quantization axis), i.e., the expectation value of the $\hat{S}^z$ operator on a given sublattice.

We introduce  bond operators, ${\hat{A}_{ij}:=\hat{a}_i\hat{b}_j-\hat{b}_i\hat{a}_j}$ and ${\hat{B}_{ij}:=\hat{a}_i\hat{b}_j+\hat{b}_i\hat{a}_j}$. Consequently, we can represent the Hamiltonian in \cref{eq:Heisenberg} in terms of these bond operators~\cite{auerb94}. In order to benefit from the system's full translational invariance, it is advantageous to rotate the spins in one sublattice, for instance, in our case sublattice $R$ with all $j$'s sites by 180$^\circ$ about the $y$ axis. The action of the rotation operator $\hat{R}_y(\pi)=\exp(-i\pi \hat{S}_j^y)$ on the state is a unitary operation, which does not change the physical properties of the system. The rotation changes the Schwinger bosons as follows 
\be\label{eq:rotation}
\begin{pmatrix}
  \hat{a}_j \\
  \hat{b}_j \\
\end{pmatrix} \rightarrow \hat{R}_y(\pi)\begin{pmatrix}
  \hat{a}_j \\
  \hat{b}_j \\
\end{pmatrix}
\hat{R}_y(\pi)^\dag
=\begin{pmatrix}
  -\hat{b}_j \\
  \hat{a}_j \\
\end{pmatrix},
\ee
i.e., $\hat{a}_j\rightarrow -\hat{b}_j$ and $\hat{b}_j\rightarrow \hat{a}_j$. As a result, the bond operators take the form ${\hat{A}_{ij}=\hat{a}_i\hat{a}_j+\hat{b}_i\hat{b}_j}$ and ${\hat{B}_{ij}=\hat{a}_i\hat{a}_j-\hat{b}_i\hat{b}_j}$. Given the constraints in \cref{eq:constr}, the Hamiltonian in \cref{eq:Heisenberg} can then be expressed in terms of the bond operators,  reading
\be
  \mathcal{\hat{H}}_0 = -\frac{1}{4}\sumnn\lrg{(1+\chi)\hat{A}_{ij}^\dagger \hat{A}_{ij}+(1-\chi)\hat{B}_{ij}^\dagger \hat{B}_{ij}-4S^2 }.
	\label{eqn:Hbond}
\ee
The exchange coupling $J$ serves as the energy unit henceforth, i.e., it is set to unity.

\subsection{The mean-field approximation}

Another advantage of the post-rotation bond operators is that the mean-field approximation becomes compactly applicable as
\bes \label{eq:MFA}
\begin{align}
   \hat{A}_{ij}^\dagger \hat{A}_{ij}\approx\langle \hat{A}_{ij}^\dagger \rangle \hat{A}_{ij}+\langle \hat{A}_{ij} \rangle \hat{A}_{ij}^\dagger-\langle \hat{A}_{ij}^\dagger \rangle \langle \hat{A}_{ij}\rangle,
\\
\hat{B}_{ij}^\dagger \hat{B}_{ij}\approx\langle \hat{B}_{ij}^\dagger \rangle \hat{B}_{ij}+\langle \hat{B}_{ij} \rangle \hat{B}_{ij}^\dagger-\langle \hat{B}_{ij}^\dagger \rangle \langle \hat{B}_{ij}\rangle. 
\end{align}
\ees
Assigning 
${A^*:=\langle \hat{A}_{ij}^\dagger \rangle}$, ${A:=\langle \hat{A}_{ij} \rangle}$  and ${B^*:=\langle \hat{B}_{ij}^\dagger \rangle}$, ${B:=\langle \hat{B}_{ij} \rangle}$ and considering the restriction on the number of the Schwinger bosons on  each lattice site by Lagrange terms (the same ${\lambda_i=\lambda_j=\lambda}$ for each lattice site) we obtain
\begin{align}
\nonumber
     \mathcal{\hat{H}}_\text{MF}& = E_\text{MF}-\frac{1}{4}\sumnn\Big(C_-^*\nbij{a}+C_-\nbdij{a}+C_+^*\nbij{b}+C_+\nbdij{b}\Big) 
     \\
  &+\lambda \sum_{i\in R}(\nbi{a}+\nbi{b})+\lambda \sum_{j\in R'}(\nbj{a}+\nbj{b}),
  \label{eq:Hblin}
\end{align}
where ${E_\text{MF}=(zN_R/4)\Big((1+\chi)A^*A+(1-\chi)B^*B+4S^2\Big)-4N_RS\lambda }$ and  $N_R=N_{R'}$, $z$ the coordination number, and $C_\pm$  contains the mean-field parameters:  ${C_{\mp}=A(1+\chi)\pm B(1-\chi)}$ together with its complex conjugate ${C_{\mp}^*=A^*(1+\chi)\pm B^*(1-\chi)}$. It is our standard convention to employ the minus index, e.g., $C_-$ for the 
variables of the $a$ bosons, and the plus index, e.g., $C_+$ for the variables of the $b$ bosons, 
respectively~\cite{bolsm23}.  
We stress that $A$ and $B$ are mean-field averages (scalars) and do not have any dependence on the lattice sites.

\subsection{Symmetry breaking auxiliary field}

We construct self-consistent mean-field equations to find mean-field ans\"atze and compute solutions for the self-consistency conditions numerically for large, but finite clusters. These solutions will serve as the initialization for switching processes.  However,  the spontaneous symmetry breaking is precluded due to the finite size of the clusters. The lowest energy state of the system is still degenerate ( $\mathbb{Z}_2$ symmetry) along $z$ for the anisotropic system under consideration.  Consequently, the rigorous determination of finite sublattice magnetization is not possible. To obtain a proper ground state of the system and the sublattice magnetization, we include  a staggered Zeeman term with a very weak auxiliary external magnetic field that couples to spins
\begin{equation}
    \mathcal{\hat{H}}_\text{m}=-h^z\lr{\sum_{i\in R} \hat{S}^z_i-\sum_{j\in R'} \hat{S}^z_j}.
\end{equation}
Because of the sublattice rotation on one lattice site in \cref{eq:rotation} the  field along the $z$ axis becomes uniform. 
Then, the aforementioned Zeeman term in Schwinger boson language reads
\begin{equation}\label{eq:alternatH}
    \mathcal{\hat{H}}_\text{m} =-\frac{h^z}{2}\sum_{i\in R}\lr{\nbi{a} - \nbi{b}}-\frac{h^z}{2}\sum_{j\in R'}\lr{\nbj{a} - \nbj{b}}.
\end{equation}
We choose the value of the $h^z$ field such that the isotropic finite-size honeycomb lattice acquires the sublattice magnetization of the infinite size system~\cite{oitmaa92,richter2004}. We provide the discussion of the auxiliary field in \cref{app:hz}. 
This auxiliary field is switched off for the actual switching process. Since our simulations consider a large cluster of
$2\times500\times500$ sites, the strength of $h^z$ is of the order of inverse the system size $N$ (${h^z\rightarrow 0}$ 
in the limit ${N \rightarrow \infty}$ ) and does not affect the switching dynamics themselves.

\subsection{Diagonalization of the mean-field Hamiltonian}

The Schwinger bosons are Fourier transformed independently on each sublattice to write the Hamiltonian in the momentum basis 
\begin{subequations}
\label{eqn:fourier}
\begin{align}
\hat{a}_i= N_R^{-1/2}\sumk e^{-i\mathbf{k}\cdot\mathbf{r}_i}\hat{a}_\mathbf{k}^R, & \quad
\hat{b}_i = N_R^{-1/2} \sumk e^{-i\mathbf{k}\cdot\mathbf{r}_i}\hat{b}_\mathbf{k}^R,
\\
\hat{a}_j = N_{R'}^{-1/2}\sumk e^{-i\mathbf{q}\cdot\mathbf{r}_j}\hat{a}_\mathbf{q}^{R'}, & \quad
\hat{b}_j = N_{R'}^{-1/2}\sumk e^{-i\mathbf{q}\cdot\mathbf{r}_j}\hat{b}_\mathbf{q}^{R'},
\end{align}
\end{subequations}
in accordance with Ref.~\citen{bena_RemarksTightbindingModel_2009}.
Using the above Fourier transformations, the mean-field Hamiltonian in momentum space reads
\begin{align} \nonumber
    \mathcal{\hat{H}}_\mathrm{MF}&=E_\mathrm{MF}+\Big(\lambda-\frac{h^z}{2}\Big)\sumk\lr{\hat{a}_\kk^{R\dagger}
    \hat{a}_\kk^R+\hat{a}_\kk^{R'\dagger}\hat{a}_\kk^{R'}}
    \\ \nonumber
    +&\Big(\lambda+\frac{h^z}{2}\Big)\sumk\lr{\hat{b}_\kk^{R\dagger}\hat{b}_\kk^R+\hat{b}_\kk^{R'\dagger}\hat{b}_\kk^{R'}}
    \\ \nonumber
    -&\frac{z}{4}\sumk\Big(C_-^*\gam^R \hat{a}_\kk^R\hat{a}_{-\kk}^{R'}+C_-\gam^{R'} \hat{a}_\kk^{R\dagger}\hat{a}_{-\kk}^{R'\dagger}
    \\
    &+C_+^*\gam^R \hat{b}_\kk^R\hat{b}_{-\kk}^{R'}+C_+\gam^{R'} \hat{b}_\kk^{R\dagger}\hat{b}_{-\kk}^{R'\dagger}\Big) , \label{eq:MFH}
\end{align}
where  $\gam^R=(1/z)\sum_l\exp(i\kk \cdot\boldsymbol{\delta}_l)$. By defining  the $\boldsymbol{\delta}_l$ 
vectors connecting the lattice sites  in \cref{fig:illus} as ${\boldsymbol{\delta}_1=(\sqrt{3}/2,{1/2})}$, \quad ${\boldsymbol{\delta}_2=(-{\sqrt{3}/2},{1/2})}$, \quad ${\boldsymbol{\delta}_3=(0,-1)}$, 
the geometrical factor $\gam^R\in \mathbb{C}$ takes the form 
\be
\gam^R=\frac{1}{3}\lr{\exp(-ik_y)+2\exp({ik_y/2})\cos{({\sqrt{3}/2}k_x)}}.
\ee
Here, we set the bond length between adjacent sites to unity.
 For the site $R'$,  $\gam^{R'}$   is obtained by  $\gam^{R'}=(\gam^{R})^*$.

To diagonalize the mean-field Hamiltonian, we use the following canonical Bogoliubov transformations~\cite{mattss94}
\bes
\label{eqn:Bogola}
 \begin{align}
        \hat{a}_\kk^R &= \exp(-i\phi_\kk^R)\cosh({\theta_\kk^R})\hat{\alpha}_\kk^R+\sinh({\theta_\kk^R})\hat{\alpha}_\kkm^{R'\dagger} ,
\\
        \hat{a}_\kk^{R'} &= \exp(-i\phi_\kk^{R'})\cosh({\theta_\kk^{R'}})\hat{\alpha}_\kk^{R'}+\sinh({\theta_\kk^{R'}})\hat{\alpha}_\kkm^{R\dagger},
\\
        \hat{b}_\kk^R &= \exp(-i\phi_\kk^R)\cosh({\theta_\kk^R})\hat{\beta}_\kk^R+\sinh({\theta_\kk^R})\hat{\beta}_\kkm^{R'\dagger} ,
\\
        \hat{b}_\kk^{R'} &= \exp(-i\phi_\kk^{R'})\cosh({\theta_\kk^{R'}})\hat{\beta}_\kk^{R'}+\sinh({\theta_\kk^{R'}})\hat{\beta}_\kkm^{R\dagger} .
\end{align}
\ees
The bosonic commutation relations require ${\phi_\kk^R=\phi_{-\kk}^{R'}}$ and ${\theta_{\kk}^R=\theta_{-\kk}^{R'}}$. 
For appropriately chosen angles, these relations diagonalize the mean-field Hamiltonian. 
Inserting the Eqs.\ \eqref{eqn:Bogola} into  \cref{eq:MFH} and using the condition that all anomalous terms vanish, i.e., pair creation and annihilation terms, as well as ${C_-\gam^{R'}\exp(i\phi_\kk^R)=C_-^*\gam^R\exp(-i\phi_\kk^R)}$ and 
${C_-\gam^R\exp(i\phi_\kk^{R'})=C_-^*\gam^{R'}\exp(-i\phi_\kk^{R'})}$, one obtains the diagonal mean-field Hamiltonian in terms of
$\alpha$ and $\beta$ bosons
\begin{align}
    \mathcal{\hat H}_\mathrm{MF}&=E_0+\sumk\omega_\kk^{-} \big(\alk^{R\dagger}\alk^R + \alk^{R'\dagger}\alk^{R'}+1\big)
    \\
    &+\sumk\omega_\kk^{+} \big(\bek^{R\dagger}\bek^R+\bek^{R'\dagger}\bek^{R'}+1\big), \nonumber
\end{align}
where $  E_0=E_\mathrm{MF}-2\lambda N_R$. Since the constant energy $E_0$  
does not affect the dynamics, we omit it from now on. The mean-field dispersions  have the form
\bes 
\label{eq:disper}
\begin{align} \label{eq:dispera}
  \omega_\kk^-:&=\omega_{\kk,\alpha}^{R}=\omega_{\kk,\alpha}^{R'}
  =\sqrt{\Big(\lambda-\frac{h^z}{2}\Big)^2-\Big(\frac{3}{4}\abs{C_-\gam^R}\Big)^2},
\\ \label{eq:disperb}
\omega_\kk^+:&=\omega_{\kk,\beta}^{R}=\omega_{\kk,\beta}^{R'}
=\sqrt{\Big(\lambda+\frac{h^z}{2}\Big)^2-\Big(\frac{3}{4}\abs{C_+\gam^R}\Big)^2}.
\end{align}
\ees
We emphasize that the modes on the one sublattice are degenerate with those of the of the other sublattice 
($\omega_{\kk,\alpha}^{R} = \omega_{\kk,\alpha}^{R'}$ and analogous for the $\beta$-mode) because there is no fundamental difference between the 
sublattices, i.e., they are linked by a reflection symmetry.


\subsection{The equilibrium state on a honeycomb lattice: system initialization}

We calculate expectation values of bilinear operators as ${\langle \mathcal{\hat{O}}\rangle=\bra{\psi}\mathcal{\hat{O}}\ket{\psi}}$, where $\ket{\psi}$ denotes the ground state of the system. First, the expressions for $a$ bosons read 
\bes
\begin{align} 
    \lara{\nbi{a}}&=\frac{1}{N_R}\sum_{i\in R}\lara{\nbi{a}}=\frac{1}{N_R}\sumk\lara{\hat a_\kk^{R\dagger}\hat a_\kk^R} 
    \\ 
    &=\frac{1}{N_R}\sumk\lr{\frac{\lambda-\frac{h^z}{2}}{2\omega_{\kk,\alpha}^{R}}-\frac{1}{2}}, \label{a1daga1}
    \\ 
     \langle \nbj{a}\rangle&=\frac{1}{N_{R'}}\sum_{j\in R'}\lara{\nbj{a}}=\frac{1}{N_{R'}}\sumk\lara{\hat a_\kk^{R'\dagger}\hat a_\kk^{R'}}
     \\
     &=\frac{1}{N_{R'}}\sumk\lr{\frac{\lambda-\frac{h^z}{2}}{2\omega_{\kk,\alpha}^{R'}}-\frac{1}{2}},
     \\ 
     \langle \nbij{a}\rangle&=\frac{1}{zN_R}\sum_{i\in R,\mathbf{\delta}}\lara{\hat a_i\hat a_{i+\mathbf{\delta}}}=\frac{1}{N_R}\sumk\gam^R\lara{\ak^R \akm^{R'}}
     \\
     &=\frac{3}{8N_R}\sumk\frac{C_-\abs{\gam^R}^2}{\omega_\kk^-}. \label{expect:a}
 \end{align}
\ees
The analogous calculations apply for $b$ bosons and we reach the following relations
\bes
\begin{align}
    \lara{\nbi{b}}&=\frac{1}{N_R}\sumk\lr{\frac{\lambda+\frac{h^z}{2}}{2\omega_{\kk,\beta}^{R}}-\frac{1}{2}} \label{b1dagb1},
    \\
     \langle \nbj{b}\rangle&=\frac{1}{N_{R'}}\sumk\lr{\frac{\lambda+\frac{h^z}{2}}{2\omega_{\kk,\beta}^{R'}}-\frac{1}{2}},
     \\
     \langle \nbij{b}\rangle&=\frac{3}{8N_R}\sumk\frac{C_+\abs{\gam^R}^2}{\omega_\kk^+}. \label{expect:b}
 \end{align}
\ees
The resulting set of equations to be solved self-consistently reads
\bes \label{eq:self}
\begin{align}
    A&=\lara{\nbij{a}}+\lara{\nbij{b}}=\frac{3}{8N}\sumk \abs{\gam^R}^2\lr{\frac{C_-}{\omega_\kk^-}+\frac{C_+}{\omega_\kk^+}}, \label{eq:self1}
    \\
    B&=\lara{\nbij{a}}-\lara{\nbij{b}}=\frac{3}{8N}\sumk \abs{\gam^R}^2\lr{\frac{C_-}{\omega_\kk^-}-\frac{C_+}{\omega_\kk^+}},  \label{eq:self2}
    \\
    4S&=\lara{\nbi{a}}+\lara{\nbi{b}}+\lara{\nbj{a}}+\lara{\nbj{b}} \nonumber
    \\
    &=\frac{2}{N}\sumk\lr{\frac{\lambda-\frac{h^z}{2}}{2\omega_\kk^-}+\frac{\lambda+\frac{h^z}{2}}{2\omega_\kk^+}-1},  \label{eq:self3}
\end{align}
\ees
where $N:=N_R=N_{R'}$. Note that the constraints in Eqs.~\eqref{eq:constr} are imposed on average. The solution to these equations  suffices to initiate the system with the appropriate mean-field parameters $A \in \mathbb{C}$, $B \in \mathbb{C}$ and $\lambda \in \mathbb{R}$ for the given anisotropy parameter $\chi$. The construction of the unit cell in the honeycomb lattice and the discretization of the lattice momentum in the Brillouin zone are given in \cref{app:unitcell}. 

The sublattice magnetization is calculated by
\be 
\label{eq:mag}
m=\frac{1}{N}\sumk\lr{\lara{\hat a_\kk^{R\dagger}\hat a_\kk^R}-\lara{\hat b_\kk^{R\dagger}\hat b_\kk^R}}.
\ee
Alternatively, the calculation of the sublattice magnetization of site $R'$ in \cref{eq:magnetization} can be performed in a similar manner. 
In order to initiate the process from a state of positive magnetization, the system is initialized such that the number of $a$ bosons on the sublattice exceeds the number of $b$ bosons. This is achieved by tuning $h^z$ to  a positive, but small value; for
details and the analyses of the dispersion see \cref{app:disp}.  

\subsection{Continuous Similarity Transformation}
In the equilibrium case we also utilize the approach sketched in Refs.~\citen{Powalski2015,Powalski2018,Walther2023,Hering2024}.
Starting from the same spin Hamiltonian of \cref{eq:Heisenberg}, we derive a self-consistent mean-field solution in the thermodynamic limit using the Dyson-Maleev representation instead of the Schwinger boson representation.
Subsequently, we transform the resulting Hamiltonian with \gls{cst} to incorporate magnon-magnon interactions which goes beyond the usual linear spin-wave theory treatment.
Concerning the first step, we begin with the classical N\'eel state.
Employing the Dyson-Maleev transformation \cite{Dyson_1956,malee58b}, we obtain a bosonic Hamiltonian that is inherently non-Hermitian.
A mean-field decoupling \cite{Takahashi_1989,Hida_1990}, a Fourier transformation, and a Bogoliubov transformation are carried out subsequently to obtain a nonlinear spin-wave theory Hamiltonian. 
Compared to previous applications of the method we refrain from solving the Bogoliubov transformation in  a self-consistent fashion, as a self-consistent solution is not obtainable for the full range of $\chi$.  
At this point it is already possible to extract results on a mean-field level, which we refer to as \gls{dmmft}.

Due to the Dyson-Maleev transformation and its resulting non-Hermitian
Hamiltonian, we employ a \gls{cst} rather than the more
commonly used continuous unitary transformation \cite{Wegner_1994,Mielke_1998,Knetter_2000,Knetter_2003}.
However, the fundamental concept remains the same, despite this difference.
A simpler, more manageable effective Hamiltonian $\mathcal{\hat{H}}_\text{eff}$ is reached by continuously transforming the initial quantum many-body Hamiltonian $\mathcal{\hat{H}}_0$.
A detailed description of this method is beyond the scope of this work but can be found in 
Refs.~\cite{Walther2023,Hering2024} and with a special focus on the honeycomb lattice  in Ref.~\cite{Kramer2025}.
A key element of the \gls{cst} is that the form of the effective Hamiltonian $\mathcal{\hat{H}}_\text{eff}$ can be altered by using different generators.
In this work, we focus on the \gls{0n} generator which disentangles 
the ground state from all higher quasi-particle sectors in the effective Hamiltonian $\mathcal{\hat{H}}_\text{eff}$ . 
There exists also a more ambitious generator, the \gls{qpc} generator, which tries to decouple all quasi-particle sectors from one another.
But the flow induced by the \gls{qpc} generator diverges for $\chi>\num{0.575+-0.005}$, 
which we attribute to an overlap of the single-magnon subspace with the three-magnon subspace beyond this threshold.
This is in accordance with prior studies performed in 
the Heisenberg limit $\chi=1$, observing a vanishing single-magnon weight at certain points of the Brillouin zone boundary~\cite{Kramer2025}.
To obtain results for the full parameter range we refrain form the \gls{qpc} generator and stick with the \gls{0n} generator. 

It is important to note that after the \gls{0n} flow, the single-magnon subspace is still connected to the
three-magnon subspace. Therefore, single-magnon properties  are determined approximately via a diagonalization in the direct sum of the single- and three-magnon subspace. This procedure is only an approximation since the three-magnon subspace is in turn connected to higher-magnon subspaces in the effective Hamiltonian.

\section{Equilibrium state analysis} 
\label{sec:equil}

\subsection{The system initialization}
First, we solve self-consistency \cref{eq:self1,eq:self2,eq:self3} and define the mean-field parameters. In order to ensure that we start from the correct initial state, we analyze the initial properties of the honeycomb lattice in terms of the Schwinger boson mean-field approach, i.e., the
equilibrium properties. At zero temperature,
the overall, collective behavior of antiferromagnets is well reproduced~\cite{oitmaa92, weih91,kados24}.  To show this, the results are compared with the results of other approximate approaches and numerical methods to justify its applicability as well.  
We shift the discussion of the auxiliary field to \cref{app:hz} where it is  demonstrated that ${h^z=0.3964 \,JN^{-1}}$ 
is a suitable value for the initialization of the symmetry broken considered cluster of ${2\times500\times500}$ sites.

\subsection{The spin gap} \label{app:spingap}

Switching corresponds to transferring  bosons from one bosonic branch to the other. The system possesses an energy barrier due to its
spin anisotropy that must be overcome to achieve such transferral. In parallel, this energy barrier also provides stability 
to the magnetic order, protecting the system from unintentional change of the orientation of the magnetic order due to external noise
or other imperfections.  It turns out that at zero temperature this energy barrier is fairly accurately given by 
the spin gap \cite{bolsm23}, see also below, i.e., the minimum energy required to excite a magnon \cite{auerb94}. 
In the Schwinger boson language, a magnon  corresponds to a composite excitation, i.e., creation of one bosonic flavor requiring energy and annihilation of the other flavor yielding energy. Hence, in the mean-field approach the spin gap is given by
\be
\Delta:=\Delta^+-\Delta^-,
\ee
where $\Delta^+:=\omega^+|_\mathbf{k=0}$ and $\Delta^-:=\omega^-|_\mathbf{k=0}$.

\begin{figure}
    \includegraphics[width=\linewidth]{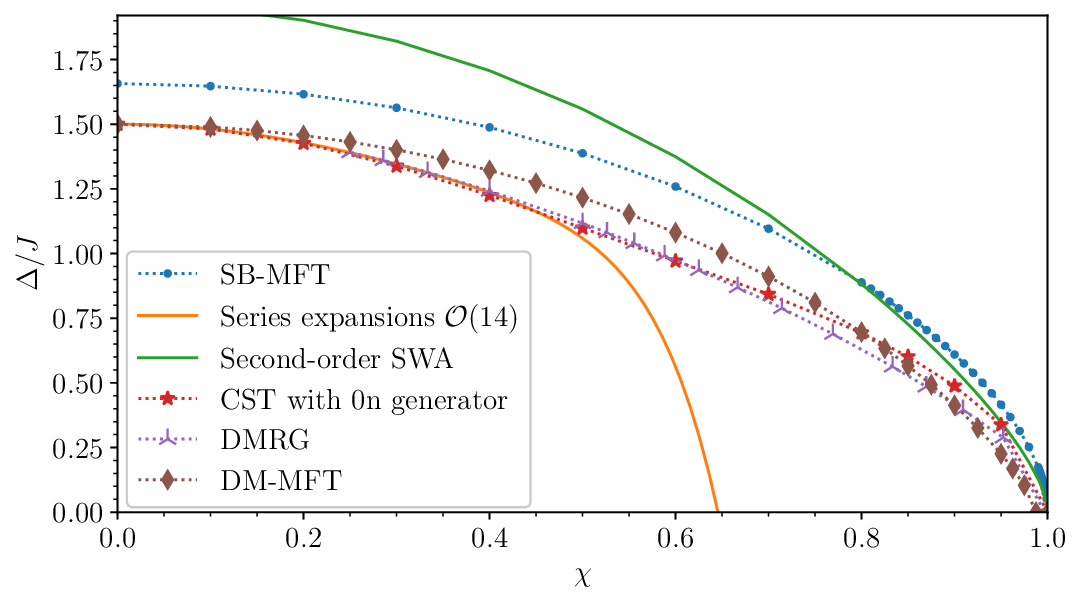}
    \caption{The spin gap vs. the anisotropy parameter.  The blue circles with dotted line are obtained by the \gls{sbmft} as employed in this article. The orange line displays results of the plain series expansions around the Ising limit from Oitmaa $et \, al$. in Ref.~\citen{oitmaa92}. The green line is the result of the second-order spin-wave analysis in Ref.~\citen{weih91}. The red stars correspond to the \gls{cst} \gls{0n} generator. The diamond shape markers are obtained by \gls{dmmft}. The tri symbols belong to the \gls{dmrg} method from Ref.~\citen{kados24}.}
    \label{fig:gap}
\end{figure}

We analyze the dependence of the spin gap on the anisotropy parameter. The mean-field theory of Schwinger bosons provides an approximate 
description of the spin gap \cite{auerb94}. 
To support this claim, we compare the results with those of other methods.
The spin gap dependence on the anisotropy is shown in \cref{fig:gap}.  
One can see that the results of the Schwinger boson approach yields slightly higher values compared to the results of other methods, 
such as the \gls{dmrg}~\cite{kados24}, \gls{cst} and the \gls{dmmft}. 
However, it is very advantageous that the self-consistent equations of the \gls{sbmft} converge for the whole region of the anisotropies,
whereas a series expansion around the Ising limit~\cite{oitmaa92} and \gls{dmmft} fail to describe the spin gap  for all $\chi\in[0,1]$.

We extract the spin gap from the effective Hamiltonian  after the \gls{0n} flow, then diagonalize  the single- and three-magnon subspace at  total momentum of $\mathbf{k=0}$.
Note that this is only an approximation, as the three-magnon subspace is in turn connected to higher-magnon subspaces in the effective Hamiltonian.
In this way we determine the spin gap for different linear systems sizes up to $L=18$ and use a linear extrapolation in $1/L$ to assess the thermodynamic limit.
Comparing the \gls{cst} results for the spin gap in \cref{fig:gap}, we find
good agreement with \gls{dmrg} results~\cite{kados24}.
Therefore, we conclude that the approximations used for the \gls{0n} generator has a negligible influence on the spin gap.
To estimate the effect of the flow, \cref{fig:gap} also shows the spin gap before the flow labeled as \gls{dmmft}. 
Compared to the \gls{cst} and \gls{dmrg} results the \gls{dmmft} spin gap shows some deviations. 
Thus, the additional \gls{cst} is necessary to capture the spin gap correctly. In comparison to the \gls{sbmft} results we see
that they do not fully capture the exact spin gap, but the results are reasonably close to justify and to proceed with the
\gls{sbmft} approach in non-equilibrium where the other approaches are far too demanding to be employed.


\subsection{The sublattice magnetization in the equilibrium} 
\label{app:mag}

\begin{figure}
    \includegraphics[width=\linewidth]{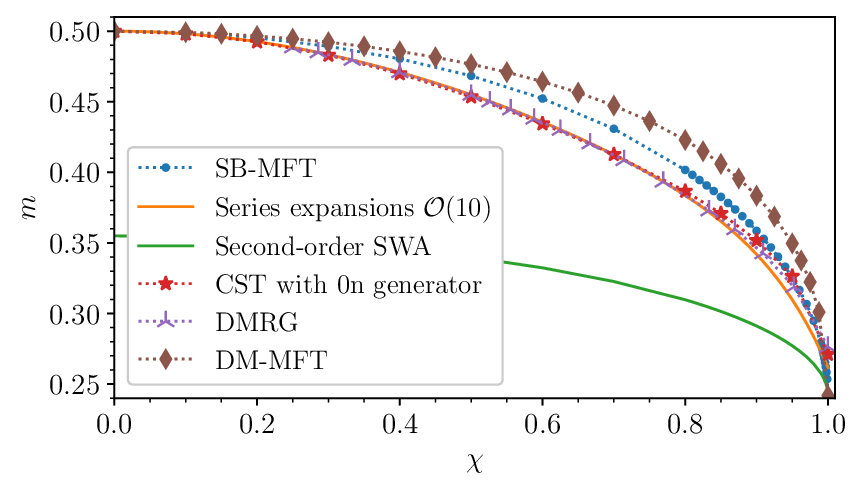}
    \caption{The sublattice magnetization dependence on the anisotropy parameter. The blue circles with dotted line depict the results of the \gls{sbmft}. The orange line is generated from the results of the series expantions around the Ising limit from Oitmaa $et \, al$. in Ref.~\citen{oitmaa92}. The green line is the result of the second-order spin-wave analysis in Ref.~\citen{weih91}. The red stars correspond to the \gls{cst} \gls{0n} generator. The diamond shape markers are obtained by the \gls{dmmft}. The tri symbols belong to the \gls{dmrg} method from Ref.~\citen{kados24}. It should be noted that the magnetization is always finite, even for $\chi=1$.}
    \label{fig:submag}
\end{figure}

The analysis of the sublattice magnetization underlines the strength of the Schwinger boson representation, even though its mean-field treatment still represents an approximation. The Schwinger boson approach functions in the entire anisotropy range in the XXZ model because the self-consistency equations converge in the whole interval of ${\chi\in[0,1]}$. The sublattice magnetization is calculated via \cref{eq:mag}. The explicit results are displayed in \cref{fig:submag} (blue circles with dotted line) as a function of $\chi$ for $S$ = 1/2. As expected, the magnetization approaches its maximum value, $S$, for ${\chi \rightarrow 0}$, at which point the model becomes an Ising model, and the limit ${\chi\rightarrow 1}$ yields
a convincing approximation of the  sublattice magnetization in the isotropic system. To show the advantage of the \gls{sbmft}, we also plotted the results of the other approaches. The magnetization curve agrees well with the results of the \gls{dmrg}~\cite{kados24} and \gls{cst}.
Here, the \gls{cst} results are determined from the continuously transformed the observable corresponding to the sublattice magnetization
\be
    \hat{M}^z = \sum_{i\in R} \hat{S}^z_i - \sum_{j\in R'} \hat{S}^z_j \quad\text{with}\quad m=\left\langle\hat{M}^z\right\rangle_0,
\ee
with the help of the \gls{cst}.
Similar to the spin gap, we use a linear extrapolation of the sublattice magnetization in $1/L$ to determine it in the thermodynamic limit.
We find the \gls{cst} results to be in perfect agreement with the \gls{dmrg} results.
Interestingly, the \gls{dmmft} results before the flow show stronger deviations compared to the \gls{sbmft} results,
indicating that the \gls{sbmft} is superior to the \gls{dmmft} in capturing the sublattice magnetization on a mean-field level.

The orange line is taken from a series expansion the around the Ising limit \cite{oitmaa92} using a
linked-cluster expansion. As demonstrated, the plain series does not provide reliable data for the  limit ${\chi\rightarrow 1}$. 
However, the reproduced result shows close agreement with the result of our approach  for  strong anisotropy, i.e., for small $\chi$.

\section{Switching the sublattice magnetization via an external magnetic field} 
\label{sec:switch}

The objective of this study is to invert the sublattice magnetization of honeycomb lattice, denoted by $m$, and transform it into its negative value $-m$. The claim is that this inversion can be attributed to the bit flip from 0 to 1 in storage devices. This process of flipping the antiferromagnetic order is imperative for the engineering of ultrafast data storage, provided that antiferromagnetic spintronics can operate at terahertz frequencies. 
Our results indeed demonstrate this ultrafast switching behavior.

The switching process is induced by an external magnetic field, which is incorporated via the Zeeman term as 
\be \label{eq:unifield}
 \mathcal{\hat{H}}_\text{u}=-\mathbf{h}_\text{u}\cdot\lr{\sum_{i\in R}\mathbf{\hat{S}}_i+\sum_{j\in R'}\mathbf{\hat{S}}_j},
\ee
where the index ``$\text{u}$'' stands for ``uniform'' field. In our approach, we apply the external field along the $y$ axis, i.e., ${\mathbf{h}_\mathrm{u}=(0,h_\mathrm{u},0)}$, where the corresponding Hamiltonian in Schwinger bosons reads
\be
 \mathcal{\hat{H}}_\text{u}=-\frac{h_\mathrm{u}}{2i}\sum_{i\in R}\Big(\hat{a}_i^\dagger\hat{b}_i-\hat{b}_i^\dagger\hat{a}_i \Big)-\frac{h_\mathrm{u}}{2i}\sum_{j\in R'}\Big(\hat{a}_j^\dagger\hat{b}_j-\hat{b}_j^\dagger\hat{a}_j \Big).
\ee
From the standpoint of quantum physics, the Fourier transform of the aforementioned Hamiltonian does not commute with the mean-field Hamiltonian in \cref{eq:MFH}. Furthermore, it incorporates bilinear bosonic operators, which contain the annihilation of one bosonic flavor and the creation of the other bosonic flavor. This is precisely what is needed in the switching process; see \cref{eq:magnetization}. The analyses commence in the state where a greater number of $a$ bosons is present. Subsequently, the application of the uniform field gives rise to an increased number of $b$ bosons, a phenomenon that occurs as a result of the annihilation of $a$ bosons.  The total number of bosons remains constant fulfilling the constraint imposed in \cref{eq:constr}.

We analyze the non-equilibrium switching of the system described by the full mean-ﬁeld Hamiltonian 
\begin{align} \nonumber
    \mathcal{\hat{H}}=&\lambda\sumk\lr{\hat{a}_\kk^{R\dagger}\hat{a}_\kk^R+\hat{a}_\kk^{R'\dagger}\hat{a}_\kk^{R'}+\hat{b}_\kk^{R\dagger}\hat{b}_\kk^R+\hat{b}_\kk^{R'\dagger}\hat{b}_\kk^{R'}}
    \\ \nonumber
    &-\frac{3}{4}\sumk\Big(C_-^*\gam^R \hat{a}_\kk^R\hat{a}_{-\kk}^{R'}+C_-\gam^{R'} \hat{a}_\kk^{R\dagger}\hat{a}_{-\kk}^{R'\dagger}
    \\ \nonumber
    &+C_+^*\gam^R \hat{b}_\kk^R\hat{b}_{-\kk}^{R'}+C_+\gam^{R'} \hat{b}_\kk^{R\dagger}\hat{b}_{-\kk}^{R'\dagger}\Big)
    \\
    &-\frac{h_\mathrm{u}}{2i}\sumk\lr{\hat a_\kk^{R\dagger}\hat b_\kk^R -\hat b_\kk^{R\dagger}\hat a_\kk^R +\hat a_\kk^{R'\dagger}\hat b_\kk^{R'}-\hat b_\kk^{R'\dagger}\hat a_\kk^{R'} }, \label{eq:MFHfull}
\end{align}
where the time-dependence of the expectation values of bilinear operators are computed using the Heisenberg equation of motion: $\partial_t\lara{\mathcal{\hat{O}}(t)}=i\lara{[\mathcal{\hat{H}},\mathcal{\hat{O}}(t)]}$ (\cref{app:EoM}). 
The auxiliary $h^z$ field is turned off for switching. 

   We solve the differential \crefrange{eqn:DissEQ1u}{eqn:DissEQ6u} at each momentum $\mathbf{k}$ in the Brillouin zone and 
   compute the occupation  of the bosons. The time dependence of the expectation values implies that the mean-field parameters are time-dependent. Consequently, a temporal adjustment of the mean-field parameters is implemented at each time step while solving the differential equations. 
   This holds also true for the Lagrange multiplier $\lambda$ fixing the total number of bosons.

\section{Results and discussion of switching} \label{sec:results}

\subsection{Magnetization switching}

\begin{figure}
    \includegraphics[width=\linewidth]{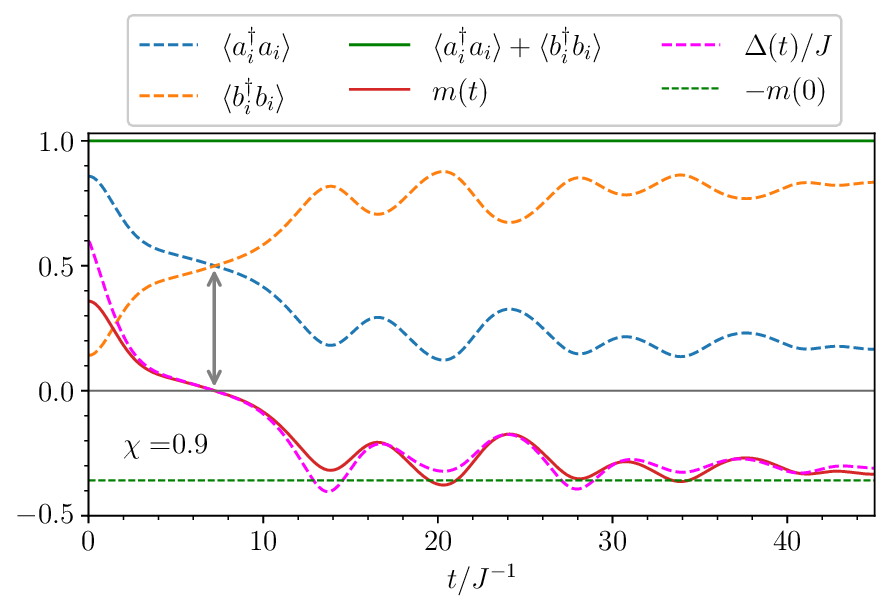}
    \caption{The dynamics of the occupations of the bosons and the magnetization. The strength of the external magnetic field is 
    $h_\mathrm{u}=0.645 \, J$. The magenta colored dashed line shows the dynamics of the spin gap. The double-headed arrow shows the instant of the switching, i.e., the sign change of $m(t)$. For comparison, the dashed green horizontal line is included to indicate the negative 
    initial magnetization.}
    \label{fig:dynamy}
\end{figure}

Figure \ref{fig:dynamy} shows the  result of the approach, namely successful switching of the sublattice magnetization in an anisotropic honeycomb lattice. The dynamics of the expectation values of the occupation of bosons (blue and orange dashed lines)  
correspond to the temporal evolution of the magnetization (solid red line), while the total bosonic occupation number (solid green line) remains conserved. 
One can see that the magnetization reaches almost its negative initial value and that the system maintains the reversed spin orientation because the state with -$m(t=0)$ is also a valid ground state. In other words, there are two symmetry-related N\'eel ground states; sublattice ${R \,\ket{\uparrow}/R'\,\ket{\downarrow}}$ and sublattice ${R \, \ket{\downarrow}/R' \,\ket{\uparrow}}$. 
These two states are degenerate in zero staggered field, but an energy barrier due to anisotropy separates them. 
Then, the system is not merely ``excited'' by the switching field.  Rather, it is driven over the anisotropy barrier by means of a 
sufficiently strong external magnetic field, thereby driving it to another distinct local minimum of the free energy. It is remarkable  
that the magnetization decreases in absolute value with decreasing oscillations and does not switch back to the initial 
$\ket{\uparrow}$ state, in spite of the switching field being still active.
We stress  that no relaxation is incorporated in the model so that we attribute this phenomenon to \textit{dephasing}. To elaborate further, the switching is a collective motion involving many spins and many modes. During the crossing of the energy barrier, 
many $\alpha$ and $\beta$ bosons are excited and propagate with their respective frequencies. 
Their weighted sum describes the magnetization causing intrinsic dephasing, i.e., the destructive interference of their temporal evolution. 
Thus, we see a decrease in magnitude and in the oscillations of the magnetization after switching (solid red line in \cref{fig:dynamy}).

As stated, the reorientation of the N\'eel vector in an antiferromagnet requires an  external energy input to overcome the potential barrier. 
In the XXZ Heisenberg model under consideration, the potential barrier is set by the anisotropy parameter $\chi$, which in turn determines the spin gap.
The gap undergoes modification in non-equilibrium states. As illustrated in \cref{fig:dynamy}, the dashed magenta line shows the 
evolution of the spin gap. Clearly, the gap is closed at the instant of switching (double-headed arrows in \cref{fig:dynamy}).
Subsequently,  it reemerges; we plot $\Delta^+ - \Delta^-$ which is the spin gap if the $a$ bosons are the majority bosons and the
sublattice magnetization  positive. Once the sign of the magnetization is flipped so that the $b$ bosons are the majority, the spin gap 
is given by $\Delta^- - \Delta^+$ so that the physical spin gap is best computed from the modulus $|\Delta^+ - \Delta^-|$. Still, we plot
$\Delta^+ - \Delta^-$ in \cref{fig:dynamy} because this illustrates remarkably synchronized dynamics of the gap and the magnetization.
 This finding indicates that they are dynamically coupled variables of the same collective modes. Indeed, in the Schwinger boson mean-field approach the spin gap depends on mean-field parameters which in turn depend on the bosonic occupations that also  determine the magnetization. 
 Our analyses show that the oscillations of the magnetization is essentially defined by the initial spin gap and the strength of the external field. 
 Moreover, we find that the frequency of the oscillations depend linearly on the initial spin gap with the same prefactor for
 all anisotropies; see \cref{app:thefreq} for further discussion.  

\begin{figure}
    \centering
    \includegraphics[width=\linewidth]{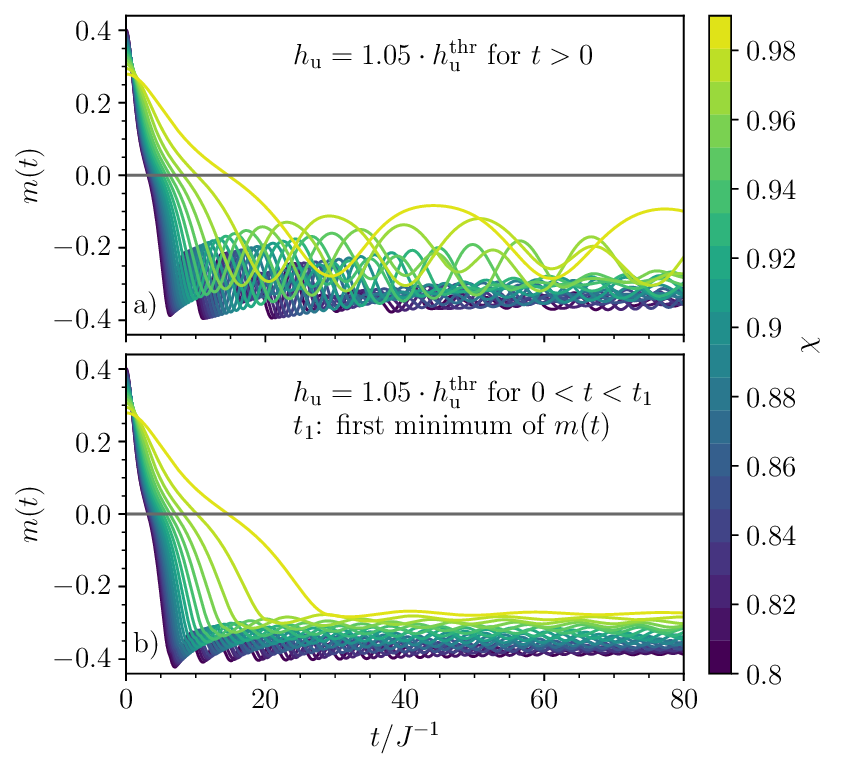}
    \caption{The dynamics of the sublattice magnetization are shown for the range of  anisotropies  $0.8\leq\chi<1$.
    Panel a) shows the dynamics for the case where the external field is applied at all times $t>0$. In panel b), the field is turned off 
    after the switching at $t_1$, where $t_1$ is the time at which $m(t)$ reaches its first minimum in the $m(t)<0$ region. 
    The minimum value of the external field for switching is labeled by the superscript ``thr'' for threshold. 
    For all anisotropies, the field strength is chosen 5$\%$ higher than the threshold value to ensure reliable switching.}
    \label{fig:magdynamicsuniform}
\end{figure}

In the simulations so far, we kept  the external field active at all times $t>0$ in \cref{fig:dynamy}. In practical applications, however, the external driving 
will be deactivated once the switching is achieved. To demonstrate the significance of this phenomenon we present the dynamics of magnetization for the 
two distinct cases in \cref{fig:magdynamicsuniform}. First, the static external field is active at all times $t>0$ for each anisotropy $\chi$
(\cref{fig:magdynamicsuniform} a)). 
The strength of the field varies for varied anisotropy based on the threshold value; see the below for more discussion on threshold values. 
One observes that the switching occurs in all $m(t)$ and the reorientation is maintained even though the switching field remains active due to dephasing.
However, the oscillations right after after switching are quite strong, so unexpected back-switching could occur. To eliminate the possibility of 
such back-switching and to keep the reoriented state reliably, we turn off the field as soon as the full switching has occurred, i.e., 
the magnetization reaches its first minimum in the $m(t)<0$ region (\cref{fig:magdynamicsuniform} b)). Then, the post-switching oscillations
have much smaller amplitude. These results demonstrate the advantageous control to reach the essentially static state with reoriented order.

\begin{figure}
    \includegraphics[width=\linewidth]{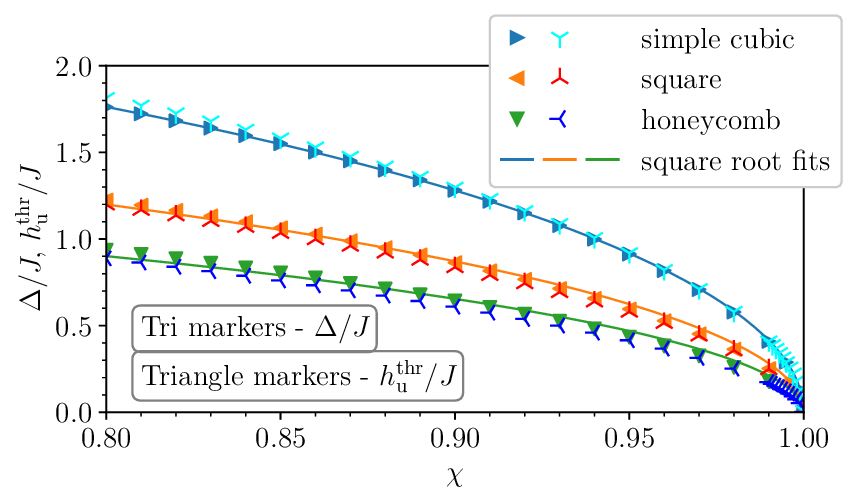}
    \caption{The uniform threshold field and the spin gap dependence on the anisotropy parameter $\chi$. 
    The triangle markers indicate the values of $h_\mathrm{u}^\mathrm{thr}$ while the tri markers show the value of the spin gap 
    $\Delta$. The solid lines are square root fits: ${h_\mathrm{u,fit}^{\mathrm{thr},\alpha}=z_\alpha c_\alpha\sqrt{1-\chi^2}}$,  $\alpha \in \{\mathrm{sc, sq, hc}\}$ 
    where $z_\alpha$ is the coordination number of the lattice and $c_\alpha$ are the fit parameters with ${c_\mathrm{sc}=(0.489 \pm 0.0002)  J}$, $z_\mathrm{sc}=6$, ${c_\mathrm{sq}=(0.499 \pm 0.0003)  J}$, $z_\mathrm{sq}=4$ and  ${c_\mathrm{hc}=(0.501 \pm 0.0044)  J}$, $z_\mathrm{hc}=3$. 
    The lattices are abbreviated by sc - simple cubic, sq - square, hc - honeycomb. The data for the square and simple cubic lattices are taken from Refs.~\citen{bolsm23,khudoy24}.}
    \label{fig:threshchi}
\end{figure}

It has been shown in hypercubic lattices that the spin gap determines the strength of the threshold field necessary to reverse the magnetization~\cite{bolsm23,khudoy24}. 
In order to corroborate this finding, the threshold values and the spin gap dependence on the anisotropy parameter are plotted for different lattices in \cref{fig:threshchi}.  One can see that the threshold fields are  almost quantitatively given by the spin gap for all lattice types. The overall square root dependence of the 
spin gap and thus the threshold field is the hallmark of mean-field treatments. Furthermore, it has been observed that the fit parameters 
$c_\alpha$ ($\alpha \in \{\mathrm{sc, sq, hc}\}$) in the fit functions ${h_\mathrm{u,fit}^{\mathrm{thr},\alpha}=z_\alpha c_\alpha\sqrt{1-\chi^2}}$ 
are very close in  value (see the caption in \cref{fig:threshchi}) while the coordination number $z_\alpha$ is the main prefactor  capturing the properties of 
the specific lattice.

We fit all threshold data simultaneously with a single function  using a common parameter in SciPy~\cite{SciPy}.  
Each function includes the coordination number $z_\alpha$ as a prefactor. To this end,  all $h_\mathrm{u}^\mathrm{thr}$
data points in  \cref{fig:threshchi} are combined and the sum of all squared residuals is minimized over the full dataset.  
The universal fit function is given by                
\be\label{eq:thrfituniform}
h_\mathrm{u}^{\mathrm{thr},\alpha}=z_\alpha c\sqrt{1-\chi^2},
\ee
where the fit parameter results to be ${c=(0.494 \pm 0.0014) J}$. 
This is one of the key findings of this article.  
We conjecture that the above formula  provides a good estimate for the minimal uniform field  required to switch the antiferromagnetic order in \textit{any}
quantum antiferromagnetic bipartite lattice with $S=1/2$ with corresponding $z_\alpha$ and $\chi$.

\subsection{Exchange-enhanced switching}

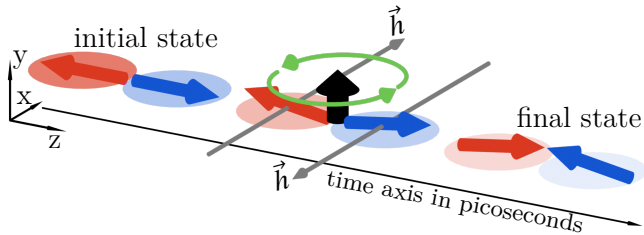
\begin{figure}
 \begin{tikzpicture}[
    every node/.style={
        font=\fontsize{11}{11}\selectfont, 
    }
]
\node at (-2.4,1.2) {initial state};
\node at (3.3,0.15) {final state};
\node at (-4.0,0.4) {x};
\node at (-3.6,-0.18) {z};
\node at (-4.05,0.9) {y};
\node [rotate=15, anchor=south west] at (-0.77,-0.99) {$\vec{h}$};
\node[rotate=15, anchor=south west] at (0.75,1.06) {$\vec{h}$};
\node[rotate=-10.5, anchor=south west,font=\small] at (-0.3,-0.9) {time axis in picoseconds};
\node at (0,0) {\includegraphics[width=\columnwidth]{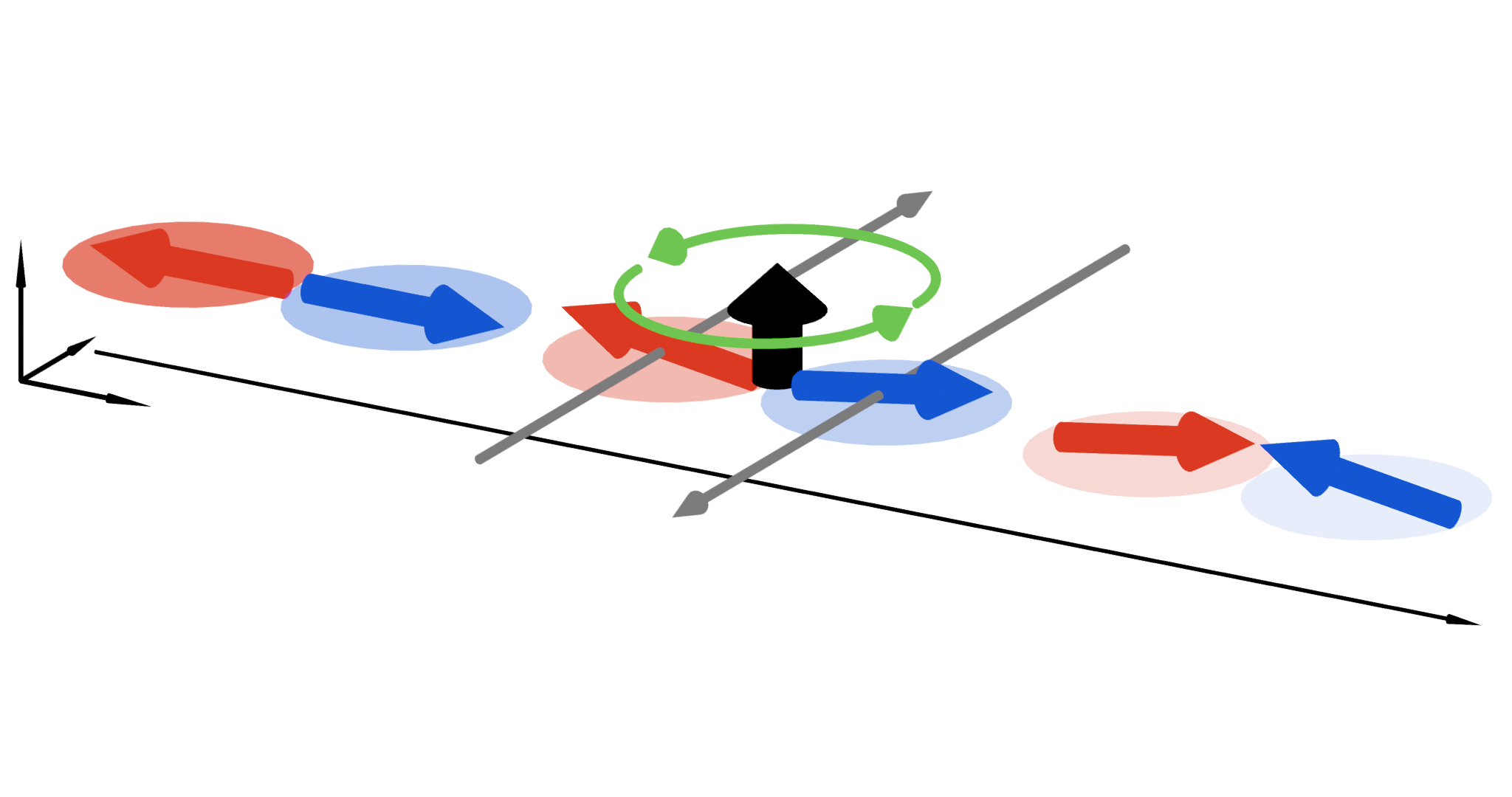}};
\end{tikzpicture}
\vspace{-1.7cm}
\caption{The classical illustration of the exchange-enhanced switching. The initial state, with two linked red and blue sublattices, is shown on the left. The arrows inside the circles indicate the direction of the sublattice magnetization. Subsequent to the application of the staggered magnetic fields on the sublattices (gray arrows), the spins exhibit a slight deflection, concurrently forming a robust effective field (black arrow). Consequently, the spins rotate around the resulting effective field (shown by the green curved arrows) and the final switched state is obtained (last two circles on the right).} \label{fig:illust_ex_switch}
\end{figure}

It has been demonstrated that a significantly weaker external \textit{staggered} field achieves switching already.
The simplified idea is that the staggered field induces spin canting in the same directions on the sublattices, where these canted 
spins generate a resulting net effective magnetic field~\cite{zelez14,khudoy25} proportional to the exchange coupling.
Then, this field induces the switching which is dubbed ``exchange-enhanced switching'', see \cref{fig:illust_ex_switch} for a detailed illustration. 
In practice, the so-called N\'eel spin-orbit torques (NSOT) realize a staggered external drive, induced by an  external global current~\cite{gomon16,wadle16}.   
In the present study, exchange enhancement is applied by replacing the switching Hamiltonian in \cref{eq:unifield}  by  
\be 
\label{eq:staggered}
 \mathcal{\hat{H}}_\text{s}=-\mathbf{h}_\text{s}\cdot\lr{\sum_{i\in R}\mathbf{\hat{S}}_i-\sum_{j\in R'}\mathbf{\hat{S}}_j},
\ee
where the index ``$\text{s}$'' stands for ``staggered'' field. The rotation around the spin $y$ axis of one sublattice in mind, 
see \cref{eq:rotation}, it is advantageous to apply the staggered field along the $x$ axis, i.e, we choose ${\mathbf{h}_\mathrm{s}=(h_\mathrm{s},0,0)}$ 
because the sublattice rotation converts it into a uniform field in the effective model. 
Further details of the calculations and the set of differential equations are provided in \cref{app:EoMexchange}.

\begin{figure}
     \includegraphics[width=\linewidth]{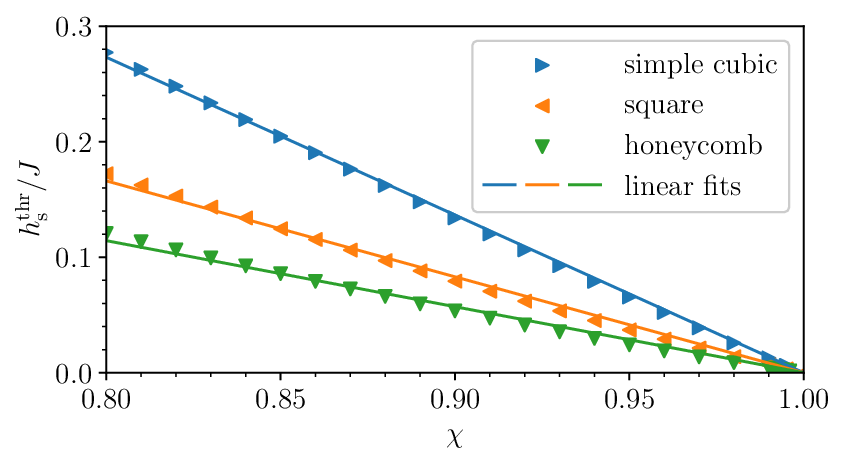}
     \caption{The staggered threshold field dependence on the anisotropy parameter $\chi$. The triangle markers correspond to the 
     $h_\mathrm{s}^\mathrm{thr}$ values while the solid lines are linear fits: 
     $h_\mathrm{s,fit}^{\mathrm{thr},\alpha}=z_\alpha d_\alpha(1-\chi)$,  $\alpha \in \{\mathrm{sc, sq, hc}\}$ where $z_\alpha$ 
     is the coordination number of the lattice and $d_\alpha$ are the fit parameters with $d_\mathrm{sc}=(0.228 \pm 0.0007)  J$ (simple cubic lattice $z_\mathrm{sc}=6$), $d_\mathrm{sq}=(0.208 \pm 0.0016)  J$ (square lattice $z_\mathrm{sq}=4$) and  $d_\mathrm{hc}=(0.191 \pm 0.0019)  J$ 
     (honeycomb lattice $z_\mathrm{hc}=3$). The data for the square and simple cubic lattices are taken from Ref.~\citen{khudoy25}. }
     \label{fig:exchange}
 \end{figure}

Indeed, the simulations of the spin dynamics on the honeycomb lattice also demonstrate that the N\'eel vector undergoes switching under 
significantly lower staggered fields in comparison to the case of uniform fields. \cref{fig:exchange} shows the threshold field dependence
on the anisotropy parameter. The observed dependence is very close to a linear relationship in contrast to the previously observed square-root behavior 
in \cref{fig:threshchi}. This is attributed to the transition in the characteristic energy from $h_\mathrm{s}$ to 
$\sqrt{Jh_\mathrm{s}}$~\cite{khudoy25,gomon16,kittel51}. The fit function ${h_\mathrm{s,fit}^{\mathrm{thr},\alpha}=z_\alpha d_\alpha(1-\chi)}$ 
contains the coordination number $z_\alpha$  and the  fit parameters $d_\alpha$.
It is noteworthy that the fit parameters are again highly similar in magnitude  for all lattices, see the caption. 
Hence, the coordination number is the primary prefactor in the linear dependence.  
Another key finding is that the field magnitudes are considerably lower than the spin gap value for all considered anisotropy regimes. 
This is indeed the justification for the claim of exchange enhancement.

Instead of fitting each dataset separately (caption in \cref{fig:exchange}), we again construct a single fit function  with a single fit parameter for all lattices, but with 
the corresponding coordination numbers of the lattices. Consequently, we obtain a single optimal fit parameter ${d=(0.216\pm  0.0018)J}$ describing
all lattices and we propose the dependence 
\be \label{eq:thrfitexchange}
h_\mathrm{s}^{\mathrm{thr},\alpha}=z_\alpha d(1-\chi),
\ee
as good universal estimate for the staggered threshold field to reorient the N\'eel vector in any antiferromagnetic bipartite lattice with $S=1/2$.  
This is another essential finding of the present study. Hence, our theory establishes that lower threshold fields are sufficient for the switching process in lattices with smaller coordination numbers. This behavior can be understood intuitively: If the spins interact with a smaller number of neighboring spins via exchange interaction,
the resulting long-range order can more easily reoriented.

\begin{figure}
    \centering
    \includegraphics[width=\linewidth]{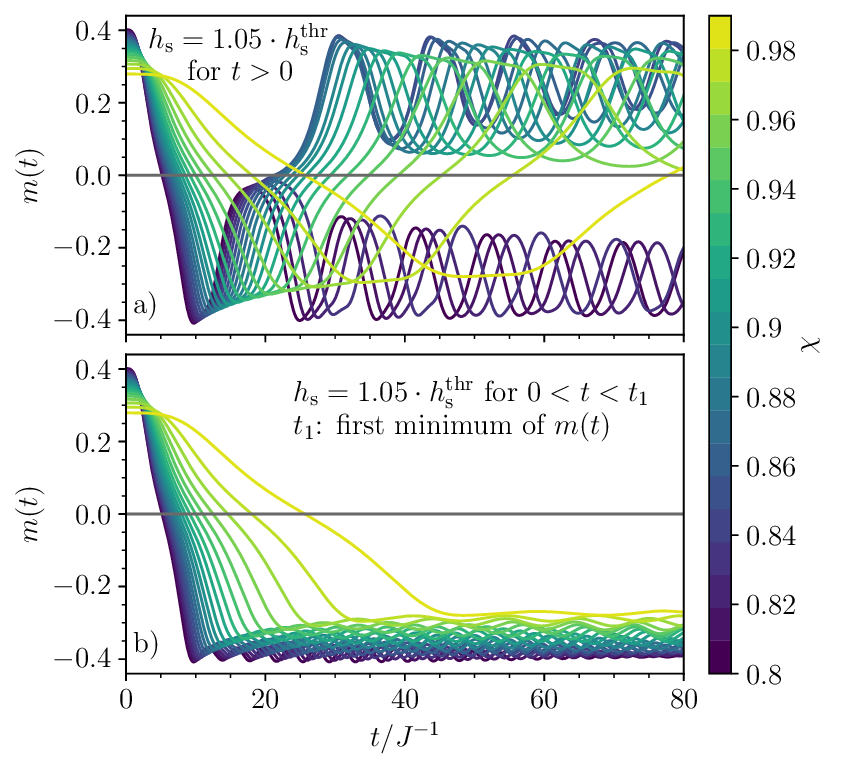}
    \caption{ The dynamics of the sublattice magnetization under staggered switching field in the anisotropy range $\chi\in[0.8,1)$. 
    Panel a) shows the dynamics when the field is active at $t>0$ whereas panel b) shows the dynamics for the case when the field is applied only 
    till the magnetization reaches its first minimum. The strength of the field is chosen always $5\%$ higher than 
    the threshold value of the field at the given anisotropy. }
    \label{fig:magdynamicsstaggered}
\end{figure}

Next, to highlight the effect of the applied field duration and of dephasing we show the dynamics of the magnetization for two cases in \cref{fig:magdynamicsstaggered},
similar to the scenario in \cref{fig:magdynamicsuniform}. First, the field is applied at all $t>0$ times (\cref{fig:magdynamicsstaggered} a)). One observes
strong oscillations of the magnetization after switching which sometimes amount up to back-switching. To avoid
detrimental back-switching we again stop the field at the first minimum of $m(t)$ in panel b). Then, the magnetization switches only once and 
subsequently exhibits damped oscillations as a consequence of dephasing (see \cref{app:thefreq} for a quantitative analysis of the oscillations after switching).

\subsubsection{Time-dependent pulses} 

In this subsection, we investigate the switching process induced by time-dependent control fields. The underlying idea is that  resonant driving should be
even more efficient. In addition, the simulation of time-dependent pulses of finite duration is much closer to experimental realizations and practical applications. 
For concreteness, we use a Gaussian shaped pulse given by the formula
\begin{align}
\label{eq:td-pulse}
h_\mathrm{s}(t)=h_\mathrm{s0}\cos(\alpha\Delta \cdot(t-3\tau)+\phi_0)\exp\left({-\frac{(t-3\tau)^2}{2\tau^2}}\right),
\end{align}
where $\alpha$ is a renormalization constant, $\phi_0$ the initial phase, $\tau$ the pulse duration and its delay: 
the pulse starts at $t=0$, but is centered at $3\tau$.  In 
\cref{fig:dynamy}, we  already saw that  the spin gap is reduced at the instant of switching indicating that the resonance does not necessarily occur at 
$\omega=\Delta$. The renormalization factor $\alpha$ in the frequency is included to account for lowered gap values.  
The pulse analyses indicate that the optimal value of the renormalization factor is $\alpha=0.85$~\cite{khudoy24}. 
The initial phase  is incorporated to ensure the optimal synchronization between the field oscillations and the system's oscillations. 
Finally, the shift ($3\tau$) of the finite duration ($\tau$) is selected to capture the essential part of the pulse; we choose $\tau =10 \,J^{-1}$
as reasonable compromise between short duration and efficient switching~\cite{khudoy24}.  

\begin{figure}
    \includegraphics[width=\linewidth]{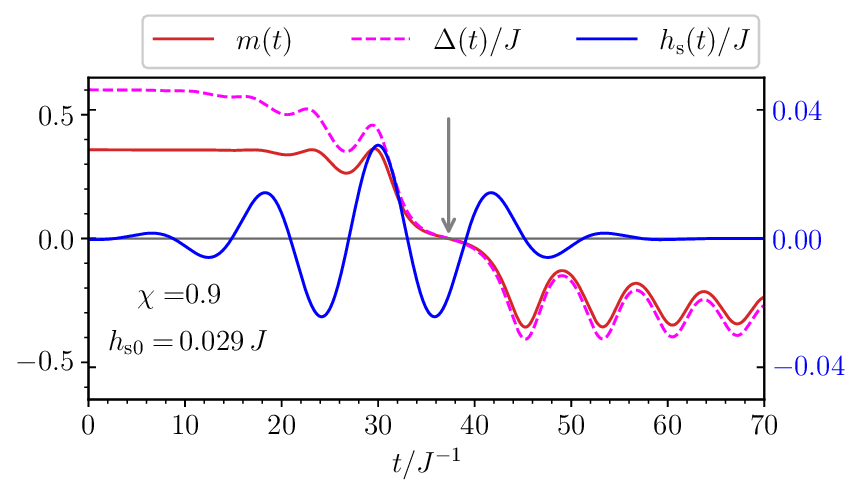}
    \caption{The dynamics of the magnetization and the spin gap under a time-dependent pulse in ~\cref{eq:td-pulse}. 
    The shape of the pulse is shown by the blue solid line and the values of pulse amplitude are given on the $y$ axis on the right-hand side.}
    \label{fig:pulse}
\end{figure}

In \cref{fig:pulse}, the dynamics of the magnetization and the spin gap are given for the pulse amplitude of $h_\mathrm{s0}=0.029\, J$. 
Remarkably, the switching occurs already under this low field whereas the threshold value of the static staggered field at $\chi=0.9$ is $
h_\mathrm{s}^\mathrm{thr}=0.053 \,J$. This is precisely the advantageous effect of resonant driving. 
As before, see \cref{fig:dynamy}, the spin gap evolves in almost perfect
synchronization to the magnetization.
 Furthermore, the dephasing effect implies decreasing oscillations in the magnetization, signifying that the system converges  to the post-switched steady-state. 
 
The local control of the sublattice magnetization using global fields is a fundamental prerequisite. The staggered fields can be realized in experiment via global fields for instance as  N\'eel-type 
spin-orbit torques induced in the sublattices by  globally applied currents~\cite{shao23}. 
Alternatively, one can use anisotropies of the $\uuline{g}$ tensors displaying differences between the sublattices. Consequently, an externally applied uniform magnetic field can generate locally staggered components. 
For the antiferromagnetic exchange coupling value of ${J=10}$ meV, the external field strength $h_\mathrm{s0}=0.029 \, J$ corresponds to about 2.5 Tesla, with the ultrafast  switching time of ${t=40 \,J^{-1}\approx 2.6 }$ picoseconds. This is already a highly promising result. Still, the anisotropy parameter $\chi=0.9$ can be regarded as strong anisotropy, i.e., the system has quite a large spin gap ($\Delta=0.609 \,J$ at $\chi=0.9$). Lower anisotropies require considerably lower either uniform or staggered fields (\cref{fig:threshchi} and \cref{fig:exchange}). Thus, in experiment the realization of the switching  with low fields is possible for weak anisotropies based on our proposal. For instance $h_\mathrm{s}=0.004\,J\approx 0.35$ Tesla is already sufficient to switch the magnetization at $\chi=0.99$.

\section{Conclusion} \label{sec:con}

The aim of the paper was to show the possibility of the N\'eel vector switching in a quantum antiferromagnetic honeycomb lattice by means of external magnetic fields. Previously, aforementioned switching has been shown in the square and simple cubic lattice using the time dependent \gls{sbmft}. 
Here, we highlighted the applicability of the theory to other lattices with lower coordination number, especially on the honeycomb lattice. 

To this end, we first analyzed the accuracy of the \gls{sbmft} in describing the equilibrium state. 
A comparison of the spin gap and the magnetization over the whole range of anisotropy revealed a reliable and satisfactory agreement with other methods,
in particular with  the \gls{cst} approach.
In spite of the bipartiteness of the lattice, the \gls{cst} approach converges for the commonly used \gls{qpc} generator 
only for strong anisotropies up to $\chi\lessapprox\num{0.575+-0.005}$. Beyond this regime, 
 divergences suggest that the single-magnon mode is no longer stable as discussed in Ref.~\cite{Kramer2025}.
To circumvent this issue the \gls{0n} generator is used and results are obtained by an additional re-diagonalization
of the one- and three-magnon subspace.
Despite this additional approximation the results agree well with \gls{sbmft} and other methods.

We demonstrated the essential physics of the switching of the sublattice magnetization 
in the quantum antiferromagnetic honeycomb lattice using the well-established Schwinger boson mean-field approach. 
We found universal formulas for the minimal value of the switching field dependence on anisotropy
where the coordination number appears as prefactor. Thus, lower threshold fields are necessary for the spin order reorientation on
lattices with smaller coordination number. This behavior is physically highly plausible 
since the robustness of spin order depends on the number of interaction partners. Hence lower coordination number imply lower
energy barriers to overcome and thus lower threshold fields.

The quantum approach  succeeds in capturing the dephasing effect in the switching process
where the superposition of all spin wave modes with different frequencies leads to destructive superposition of oscillations resulting in a slow decay. This phenomenon makes it possible to reach the steady-state of the magnetic order in the long-time limit after the switching event.
In this regard, we also highlighted  that the deactivation of the external switching field after the  reorientation of the magnetization is crucial.

The benefits of using staggered fields to reorient the magnetization are clearly demonstrated in our quantum model. 
The low staggered fields are sufficient for the reorientation of the magnetic order because of exchange enhancement. 
Specifically, despite the modest applied fields the effective torques driving the N\'eel vector are amplified by the 
large exchange field assisting the switching process.

The results of the present article pave the way for the design of state-of-the-art spintronic logic devices, 
featuring  enhanced density of data storage, robustness and ultrafast operational speed. 
The analyses of the switching on the honeycomb lattice at finite temperature and the treatment of other methods beyond the mean-field approach are
left for further investigation.  

\section*{Acknowledgment}
\begin{acknowledgment}
\acknowledgment
 We are thankful to C. Kr\"amer, K. P. Schmidt and B. Fauseweh for useful discussion and suggestions. 
 This work has been funded by the Deutsche Forschungsgemeinschaft (German Research Foundation) in project UH 90/14-2.
\end{acknowledgment}

\bibliography{reference.bib}

\appendix

\section{Lattice and reciprocal lattice vectors of the honeycomb: Discretization of the lattice momentum} \label{app:unitcell}

\begin{figure}
    \centering
    \includegraphics[width=\linewidth]{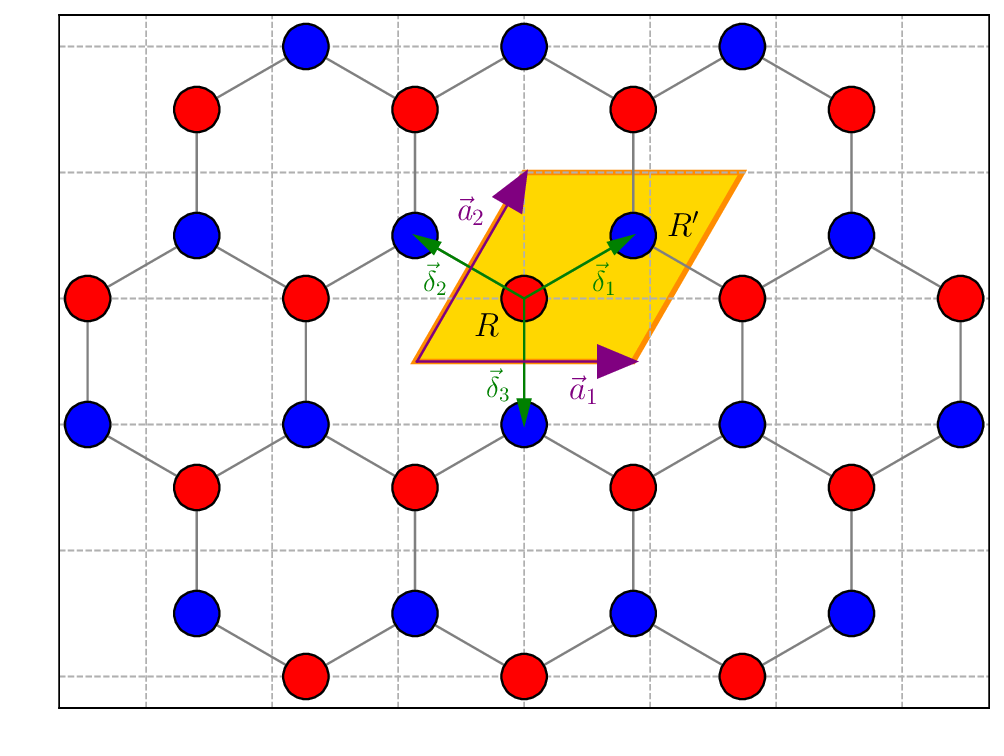}
    \caption{The unit cell of the honeycomb lattice is shown as orange shaded area. The vectors $\textbf{\textit{a}}_1$ and $\textbf{\textit{a}}_2$ are the primitive lattice vectors. }
    \label{fig:cell} 
\end{figure}

The discretization of the Brillouin zone of a honeycomb lattice requires the consideration of a two-site basis in the unit cell. 
In \cref{fig:cell}, the unit cell of the lattice is shown with corresponding basis vectors $\textbf{\textit a}_1$ and $\textbf{\textit a}_2$, 
which span the unit cell of the lattice. 
We define the lattice vectors through the basis vectors as ${\boldsymbol{\delta}_1=\frac{1}{3}(\textbf{\textit{a}}_1+\textbf{\textit{a}}_2)}$, ${\boldsymbol{\delta}_2=-\textbf{\textit{a}}_1+\boldsymbol{\delta}_1}$ and ${\boldsymbol{\delta}_3=-\textbf{\textit{a}}_2+\boldsymbol{\delta}_1}$. 
Then, we express the crystal momentum in terms of the primitive vectors $\textbf{\textit{b}}_1$ and $\textbf{\textit{b}}_2$ of the reciprocal lattice as ${\textbf{k}=m_1\textbf{\textit{b}}_1+m_2\textbf{\textit{b}}_2}$, where 
\begin{align}
    m_i\in \{-0.5, -0.5+\frac{1}{N_i}, -0.5+\frac{2}{N_i},...., 0.5-\frac{1}{N_i}\},
\end{align}
with $N_i$ being an integer, i.e., the number of unit cells in one direction. 
The total number of points is $N=N_1N_2$. In this work, the calculations are performed for a system size of $N_1=N_2=500$. 
Note that the discretization for the \gls{cst} calculation uses lower linear system sizes of $L\leq18$ with $N=L^2$ and additionally distinguishes between periodic and antiperiodic boundary conditions, for details see Ref.~\citen{Hering2024}.
The basis vectors of the lattice $\{\textbf{\textit{a}}_i\}$ are related to the basis vectors of the reciprocal lattice $\{\textbf{\textit{b}}_i\}$  through the relation ${\textbf{\textit{a}}_i\cdot \textbf{\textit{b}}_j=2\pi\delta_{ij}}$. 
Their explicit forms in Cartesian coordinates read ${\textbf{\textit{a}}_1=\lr{\sqrt{3},0}}$, ${\textbf{\textit{a}}_2=\lr{{\sqrt{3}/2},\frac{3}{2}}}$ and ${\textbf{\textit{b}}_1=\lr{{2\pi/\sqrt{3}},-\frac{2\pi}{3}}}$, ${\textbf{\textit{b}}_2=\lr{0,{4\pi/3}}}$, where the lattice spacing is set to unity. 
Now, we can compute $\gam^R$ as
\bes
\begin{align} 
   \gam^R =& \frac{1}{z} \sum_l \exp(i\kk \cdot\boldsymbol{\delta}_l) \\
   = & \frac{1}{3} \lr{\exp(i\kk \cdot\boldsymbol{\delta}_1)
   + \exp(i\kk \cdot\boldsymbol{\delta}_2) + \exp(i\kk \cdot\boldsymbol{\delta}_3) }
\\
   =& \frac{1}{3}\exp(\frac{2\pi i}{3}(m_1+m_2)) \lr{1+e^{-2\pi i m_1} + e^{(-2\pi i m_2}}.
\end{align}
\ees

\section{The initial properties of the honeycomb lattice} \label{app:init}

\subsection{The proper value of the auxiliary field} \label{app:hz}

\begin{figure}
\includegraphics[width=\linewidth]{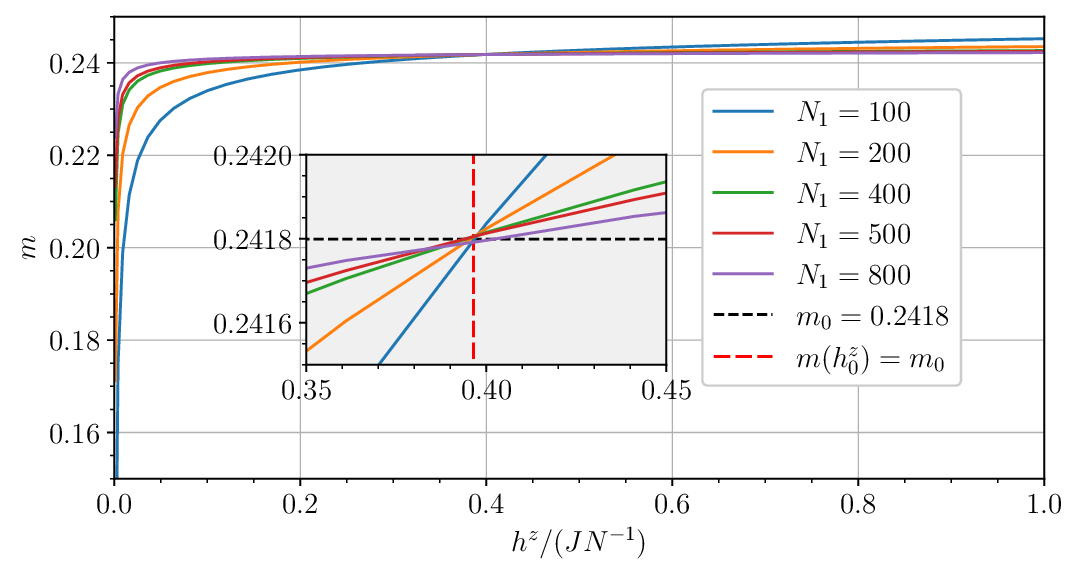}
\caption{The initial sublattice magnetization $m$ of the isotropic system ($\chi = 1$) is shown as a function of a tiny auxiliary field $h^z$ at different system sizes. The lines are interpolated to determine the value of the optimal intersection point using cubic splines from the scipy.interpolate library~\cite{SciPy}. The intersection solution is found at ${h^z_0=0.3964\,JN^{-1}}$, ${m= 0.241801}$, with a minimal variance in the $m$ values of different system sizes at this point. Note that $N=N_1N_2$ where we always consider the systems with $N_1=N_2$. A significant observation is that the magnetization curves intersect with numerical accuracy
at the analytically calculated magnetization value ($m_0=0.2418$) from Ref.~\citen{oitmaa92}, as shown in the inset. } 
\label{fig:mhz}
\end{figure}

To determine a suitable value for the auxiliary magnetic field $h^z$, the initial sublattice magnetization of the isotropic system is plotted for various system sizes for  spin $S$ = 1/2 in \cref{fig:mhz}. Notably, for all system sizes, the magnetization quickly reaches its saturated value and all lines intersect at almost a single point, as shown in the inset. We consider this value of $h^z$  as optimal. To quantify the convergence of the curves, we interpolated all curves and minimized the variance in their $m$ values.  The best common intersection point is found at ${h^z_0=0.3964\,JN^{-1}}$  with the corresponding 
magnetization value of $m=0.2418$.  The curves exhibit negligible deviation from one another at this point with the minute variance of $10^{-11}$. A remarkable result is that the defined value of the magnetization quite accurately reproduces  
the analytically determined sublattice magnetization of the honeycomb lattice,  established by Oitmaa \textit{et  al.} in Ref.~\citen{oitmaa92}. 
This fact nicely justifies our procedure to use a tiny auxiliary field. Therefore, 
we use this value of the auxiliary $h^z$ field to ensure the self-consistency equations converge properly in  equilibrium. We recall that the field is turned
off for switching and due to the substantial value of $N$ in our simulations, the actual value of $h^z$ is irrelevant and does not impact the dynamics. 

\subsection{The dispersion relations} 
\label{app:disp}

\begin{figure}
    \includegraphics[width=\linewidth]{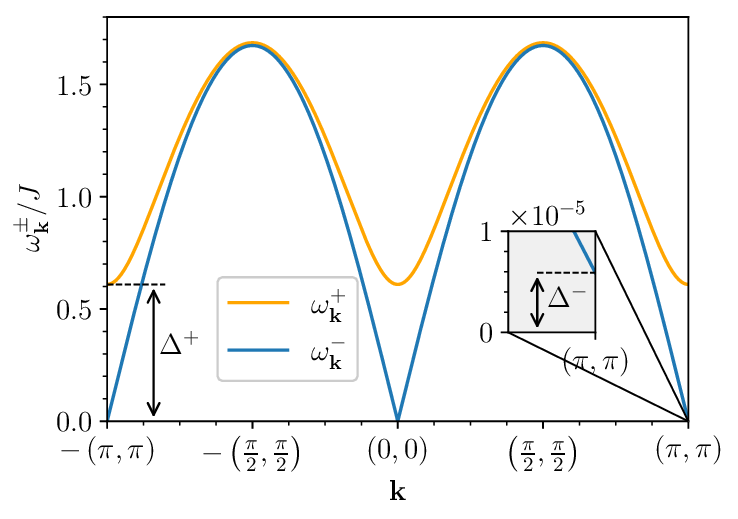}
    \caption{Dispersion of $\alpha$ bosons ($\omega_{\mathbf{k}}^-$) and $\beta$ bosons ($\omega_{\mathbf{k}}^+$) from Eqs.\ \eqref{eq:disper} at $\chi=0.9$. $\Delta^-$ and $\Delta^+$ correspond to the energy gaps of  $\alpha$ and $\beta$ bosons, respectively. The energy gap of the $\alpha$ bosons is very small 
    as shown in the inset with a value of ${\Delta^-=5.899\times 10^{-6}\, J}$. This is due to the finite size of the considered lattice. In contrast, the energy gap of the $\beta$ bosons is substantial, with a value of ${\Delta^+= 0.609 \,J}$. This indicates that the occupation of the $\alpha$ bosons is predominantly macroscopic.}
    \label{fig:disp}
\end{figure}

In \cref{fig:disp}, we provide the resulting dispersion of the two Schwinger bosons to highlight their occupation number on the two sublattices for $\chi=0.9$. Both bosonic dispersions exhibit a gap.  Actually,  for the macroscopic occupation of $\alpha$ bosons, the dispersion $\omega_\mathbf{k}^-$ should become gapless for an inﬁnitely large 
system at zero temperature. Therefore, we set the condition ${\lr{\lambda-\frac{h^z}{2}}^2=\lr{\frac{3}{4}\abs{C_-}}^2(1+\kappa)}$ in \cref{eq:dispera}, where ${\kappa \propto 1/N}$
\cite{auerb94}. This results in the formation of a minute gap in the dispersion $\omega_\mathbf{k}^-$.  To elaborate further, in our \gls{sbmft} the dispersion depends on the mean-field parameters and the Lagrange parameter. In a finite cluster, they are determined from $\mathbf{k}$-sums rather than integrals. Because the infrared contribution associated with boson condensation is cut off by ${\mathbf{k}_{min}=(\pi/N_1,\pi/N_2)}$ the mean-field values are slightly shifted away from their 
thermodynamic critical values. Consequently, the spectrum is slightly detuned and the dispersion relation for $\alpha$ bosons acquires a tiny finite-size gap (inset of \cref{fig:disp}), which vanishes in the thermodynamic limit. The small gap appears for the majority $\alpha$ bosons as we initially impose them to form the macroscopic occupation  in the equilibrium. This approach is indeed established in the literature~\cite{auerb94} for Schwinger bosons in finite-size systems.

\section{The equation of motion} \label{app:EoM}
In this Appendix, we provide the differential equations to compute the time
dependence of the expectation values, constructed by means of Heisenberg’s equations of motion.

\subsection{Switching via uniform external field}
In the case of the uniform switching field, the total mean-field Hamiltonian is given by  \cref{eq:MFHfull}. One can realize that the total mean-field Hamiltonian \eqref{eq:MFHfull} is invariant under the transformation ${(\mathbf{k},R,R')\rightarrow(-\mathbf{k},R',R)}$. This property facilitates the task of deriving and solving the closed set of differential equations. Indeed, by inserting the mean-field Hamiltonian from \cref{eq:MFHfull} into the Heisenberg equation of motion,  we obtain the following system of differential equations  
\bes
\label{eqn:DissEQu}
\begin{align}
    \partial_t \lara{\ak^{R\dagger} \ak^R} &= -i\frac{3}{4} 
     \Big(C_-^*\gam^R\lara{\ak^R\akm^{R'}}-
		C_-\gam^{R'}\lara{\ak^{R\dagger}\akm^{R'\dagger}}\Big) 
		\nonumber \\
	 & \quad + \frac{h_\mathrm{u}}{2}\Big(\lara{\ak^{R\dagger}\bk^R}+
		\lara{\bk^{R\dagger}\ak^R}\Big), \label{eqn:DissEQ1u}
\\
    \partial_t \lara{\bk^{R\dagger} \bk^R} &=-i \frac{3}{4} 
		\Big(C_+^*\gam^R\lara{\bk^R\bkm^{R'}}
		-C_+\gam^{R'}\lara{\bk^{R\dagger}\bkm^{R'\dagger}}\Big) 
		\nonumber \\ 
    &\quad - \frac{h_\mathrm{u}}{2}\big(\lara{\ak^{R\dagger}\bk^R}+
		\lara{\bk^{R\dagger}\ak^R}\big), \label{eqn:DissEQ2u}
\\
    \partial_t \lara{\ak^R \akm^{R'}} &=  i \frac{3}{4} 
		C_- \gam^{R'}\Big(\lara{\ak^{R\dagger} \ak^R}+\lara{\akm^{R'\dagger} \akm^{R'}}+1\Big)
		\nonumber \\
		&\quad - 2 \lambda  i \lara{\ak^R\akm^{R'}}_{\gamma} 
		+ \frac{h_\mathrm{u}}{2}\Big(\lara{\akm^{R'}\bk^R}+\lara{\ak^R\bkm^{R'}}\Big) ,\label{eqn:DissEQ3u}
\\
    \partial_t \lara{\bk^R \bkm^{R'}} &=  i\frac{3}{4} C_+ \gam^{R'}\Big(\lara{\bk^{R\dagger} \bk^R}+\lara{\bkm^{R'\dagger} \bkm^{R'}}+1)\Big)
		\nonumber \\
		&\quad - 2 \lambda  i \lara{\bk^R\bkm^{R'}}_{\gamma} 
		-\frac{h_\mathrm{u}}{2}\Big(\lara{\ak^R\bkm^{R'}}+\lara{\akm^{R'}\bk^R}\Big) ,\label{eqn:DissEQ4u}
\\
    \partial_t \lara{\ak^{R\dagger} \bk^R} &=  i \frac{3}{4} 
		\Big(C_+\gam^{R'}\lara{\ak^{R\dagger}\bkm^{R'\dagger}} 
		- C_-^*\gam^R\lara{\akm^{R'}\bk^R}\Big)
		\nonumber \\
		&\quad+ \frac{h_\mathrm{u}}{2}\Big(\lara{\bk^{R\dagger}\bk^R}
		-\lara{\ak^{R\dagger}\ak^R}\Big) ,\label{eqn:DissEQ5u}
\\
    \partial_t \lara{\ak^R \bkm^{R'}} &=  i \frac{3}{4}  
		\gam^{R'}\Big( C_-\lara{\akm^{R'\dagger}\bkm^{R'}} + 
		C_+\lara{\bk^{R\dagger}\ak^R} \Big)
		\nonumber \\ \nonumber
     &\quad -2 \lambda  i \lara{\ak^R\bkm^{R'}}\\
		 &\quad - \frac{h_\mathrm{u}}{2}\Big(\lara{\ak^R\akm^{R'}}-\lara{\bk^R\bkm^{R'}}\Big).
				\label{eqn:DissEQ6u}
\end{align}
\ees

\subsection{Switching via staggered field} \label{app:EoMexchange}

The Hamiltonian \eqref{eq:staggered} is written in the Schwinger boson language as
\be 
 \mathcal{\hat{H}}_\text{s}=-\frac{h_\mathrm{s}}{2}\sum_i\Big(\hat{a}_i^\dagger\hat{b}_i+\hat{b}_i^\dagger\hat{a}_i \Big)-\frac{h_\mathrm{s}}{2}\sum_j\Big(\hat{a}_j^\dagger\hat{b}_j+\hat{b}_j^\dagger\hat{a}_j \Big).
\ee
To provide the equations of motion for the expectation values in the momentum space, we rewrite the above Hamiltonian as

\begin{align} 
    \mathcal{\hat{H}}_\mathrm{s}=-\frac{h_\mathrm{s}}{2}\sumk\lr{\hat a_\kk^{R\dagger}\hat b_\kk^R +\hat b_\kk^{R\dagger}\hat a_\kk^R +\hat a_\kk^{R'\dagger}\hat b_\kk^{R'}+\hat b_\kk^{R'\dagger}\hat a_\kk^{R'} }. \label{eq:Zeemanstag}
\end{align}
Then, the equation of motion is derived through the application of the mean-field Hamiltonian, as outlined in \cref{eq:MFHfull}.
The only change is to substituting the final term by the term in  \cref{eq:Zeemanstag} to account for the change from uniform to
staggered driving field. The closed set of differential equations reads 
\bes
\label{eqn:DissEQ}
\begin{align}
    \partial_t \lara{\ak^{R\dagger} \ak^R} &= -i\frac{3}{4} 
     \Big(C_-^*\gam^R\lara{\ak^R\akm^{R'}}-
		C_-\gam^{R'}\lara{\ak^{R\dagger}\akm^{R'\dagger}}\Big) 
		\nonumber \\
	 & \quad + i\frac{h_\mathrm{s}}{2}\Big(\lara{\ak^{R\dagger}\bk^R}-
		\lara{\bk^{R\dagger}\ak^R}\Big), \label{eqn:DissEQ1}
\\
    \partial_t \lara{\bk^{R\dagger} \bk^R} &=-i \frac{3}{4} 
		\Big(C_+^*\gam^R\lara{\bk^R\bkm^{R'}}
		-C_+\gam^{R'}\lara{\bk^{R\dagger}\bkm^{R'\dagger}}\Big) 
		\nonumber \\ 
    &\quad - i\frac{h_\mathrm{s}}{2}\big(\lara{\ak^{R\dagger}\bk^R}-
		\lara{\bk^{R\dagger}\ak^R}\big), \label{eqn:DissEQ2}
\\
    \partial_t \lara{\ak^R \akm^{R'}} &=  i \frac{3}{4} 
		C_- \gam^{R'}\Big(\lara{\ak^{R\dagger} \ak^R}+\lara{\akm^{R'\dagger} \akm^{R'}}+1\Big)
		\nonumber \\
		&\quad - 2 \lambda  i \lara{\ak^R\akm^{R'}}_{\gamma} 
		+ i\frac{h_\mathrm{s}}{2}\Big(\lara{\akm^{R'}\bk^R}+\lara{\ak^R\bkm^{R'}}\Big) ,\label{eqn:DissEQ4}
\\
    \partial_t \lara{\bk^R \bkm^{R'}} &=  i\frac{3}{4} C_+ \gam^{R'}\Big(\lara{\bk^{R\dagger} \bk^R}+\lara{\bkm^{R'\dagger} \bkm^{R'}}+1\Big)
		\nonumber \\
		&\quad - 2 \lambda  i \lara{\bk^R\bkm^{R'}}_{\gamma} 
		+i\frac{h_\mathrm{s}}{2}\Big(\lara{\ak^R\bkm^{R'}}+\lara{\akm^{R'}\bk^R}\Big) ,\label{eqn:DissEQ5}
\\
    \partial_t \lara{\ak^{R\dagger} \bk^R} &=  i \frac{3}{4} 
		\Big(C_+\gam^{R'}\lara{\ak^{R\dagger}\bkm^{R'\dagger}} 
		- C_-^*\gam^R\lara{\akm^{R'}\bk^R}\Big)
		\nonumber \\
		&\quad-i \frac{h_\mathrm{s}}{2}\Big(\lara{\bk^{R\dagger}\bk^R}
		-\lara{\ak^{R\dagger}\ak^R}\Big) ,\label{eqn:DissEQ3}
\\
    \partial_t \lara{\ak^R \bkm^{R'}} &=  i \frac{3}{4}  
		\gam^{R'}\Big( C_-\lara{\akm^{R'\dagger}\bkm^{R'}} + 
		C_+ \lara{\bk^{R\dagger}\ak^R} \Big)
		\nonumber \\ \nonumber
     &\quad -2 \lambda  i \lara{\ak^R\bkm^{R'}}\\
		 &\quad + i\frac{h_\mathrm{s}}{2}\Big(\lara{\ak^R\akm^{R'}}+\lara{\bk^R\bkm^{R'}}\Big).
				\label{eqn:DissEQ6}
\end{align}
\ees

\section{The frequency of the oscillations and the spin gap} \label{app:thefreq}

\begin{figure}
    \centering
    \includegraphics[width=\linewidth]{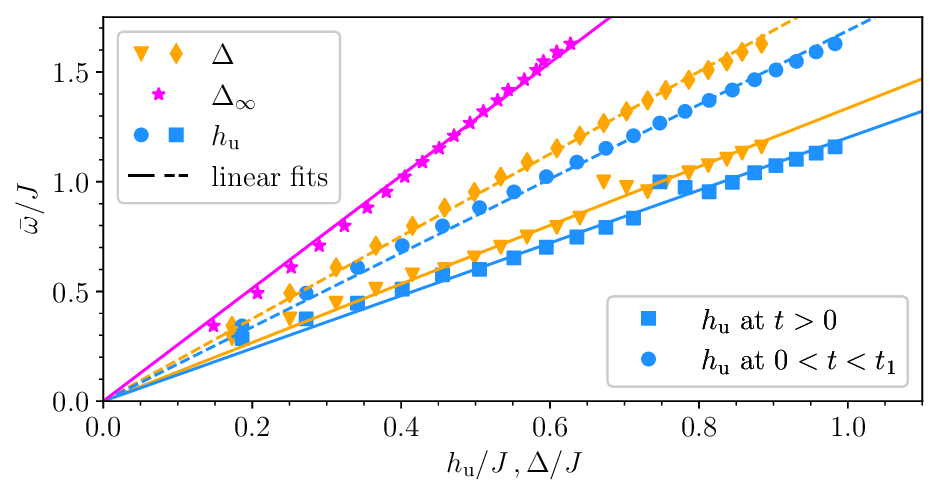}
    \caption{The frequency of the oscillations vs.\  external field and the spin gap for different anisotropies. 
    Solid blue line ${\bar{\omega}_{\mathrm{fit,1}}=k_1\cdot h_\mathrm{u}}$ where ${k_1=1.201 \pm 0.011}$. 
    Solid orange line ${\bar{\omega}_{\mathrm{fit,2}}=k_2\cdot\Delta}$, where ${k_2=1.335 \pm 0.012}$. Dashed blue line ${\bar{\omega}_{\mathrm{fit,3}}=k_3\cdot h_\mathrm{u}}$, where ${k_3=1.689 \pm 0.007}$. Dashed orange line ${\bar{\omega}_{\mathrm{fit,4}}=k_4\cdot\Delta}$, where ${k_4=1.878 \pm 0.006}$ and ${\bar{\omega}_{\mathrm{fit,5}}=k_5\cdot\Delta_\infty}$, where ${k_5=2.57 \pm 0.012}$. The time $t_1$ is the period until $m(t)$ reaches its first minimum in ${m(t)<0}$ region (see \cref{fig:magdynamicsuniform}). }
    \label{fig:freq_uni_gap}
\end{figure}

The switching occurs under strong uniform fields and thus the driving field considerably changes the initial state of the system. 
This effect prevents the system from reaching a static post-switched state 
because of large oscillations when the field is not deactivated after switching. 
To highlight this effect, we analyze the oscillations after switching for the two cases in \cref{fig:magdynamicsuniform} up to $t_{\mathrm{max}}=500 \,J^{-1}$ 
starting from the time $t_1$ and plot the average frequency vs.\ the applied fields in \cref{fig:freq_uni_gap}. The subsequent analyses reveal 
the presence of distinct frequencies in the  oscillations for the case where $h_\mathrm{u}$ is active for all $t>0$; 
see the  blue square markers and their corresponding linear fit. 
Since the uniform threshold fields closely agree with the spin gap it is not astounding that a similar linearity occurs in dependence on the spin gap,
see orange triangle markers.

The average frequency of the oscillations also show a clear linear dependence on the strength of the field (blue circle markers) and on the initial spin gap 
(the orange diamond shape markers) when the external field is deactivated after switching. 
In this case, the dephasing effect dominates and leads to strongly damped oscillations.  
The linear fit ${\bar{\omega}_{\mathrm{fit,4}}=k_4\cdot\Delta}$ results in the prefactor of ${k_4=1.878\pm 0.0065}$.

Moreover, we define the average gap $\Delta_\infty$
after switching because the system relaxes to a  steady-state with the gap which is not a ground state.  
This asymptotic long-time value of the gap $\Delta_\infty$ is defined by
\be \label{eq:gapinfinity}
\Delta_\infty=\lim_{T\rightarrow\infty}\frac{1}{T}\int_{t_1}^{t_1+T}\Delta(t)dt,
\ee 
where ${T=t_{\mathrm{max}}-t_1}$. 
In \cref{fig:freq_uni_gap}, the magenta star markers show the average frequency dependence 
on the $\Delta_\infty$ with its linear dependence, fitted by the respective line with the prefector is ${k_5=2.57\pm 0.012}$.  
These observations indicate that the frequency is approximately given by  ${\bar{\omega}\approx 2  \Delta}$. 
Indeed, this is consistent with our description including both transverse and longitudinal fluctuations of the N\'eel vector
given by  transverse  and longitudinal  magnonic modes. 
The observed oscillation frequency being approximately twice the spin gap 
can be understood as arising from the coupling of the order parameter dynamics to pairs of these gapped magnons.

\begin{figure}
    \centering
    \includegraphics[width=\linewidth]{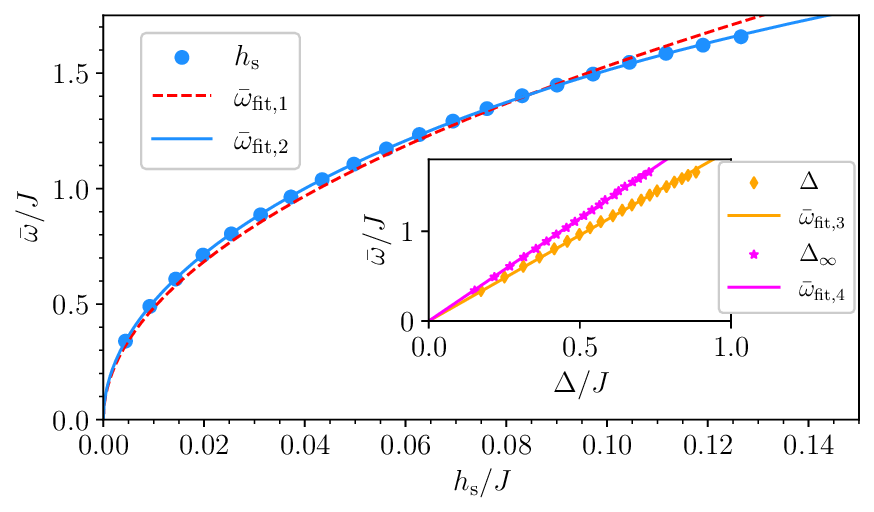}
    \caption{ The average frequency of the oscillations vs.\  the externally applied staggered field 
    in the interval ${0<t<t_1}$ based on \cref{fig:magdynamicsstaggered}b). The fit functions are: ${\bar{\omega}_{\mathrm{fit,1}}=p_1\cdot\sqrt{Jh_\mathrm{s}}}$ with ${p_1=4.8397 \pm 0.027}$. ${\bar{\omega}_{\mathrm{fit,2}}=p_2\cdot\sqrt{J(h_\mathrm{s}-d_2\cdot h_\mathrm{s}^2)}}$ with ${p_2=5.144 \pm 0.008}$ and ${d_2=(1.356 \pm 0.027)J^{-1}}$. The inset shows the frequency dependence on the spin gap with its corresponding linear fits;  ${\bar{\omega}_{\mathrm{fit,3}}=p_3\cdot\Delta}$ with ${p_3=1.912 \pm 0.005}$ and ${\bar{\omega}_{\mathrm{fit,4}}=p_4\cdot\Delta_\infty}$ with ${p_4=2.287 \pm 0.003}$.}
    \label{fig:freq_stag_gap}
\end{figure}

The behavior of the average frequency of the post-switching oscillations vs.\  $h_\mathrm{s}$ is different for exchange-enhanced switching; 
see \cref{fig:freq_stag_gap}. Note that the field is only active in the interval ${0<t<t_1}$, where $t_1$ is the time at which
the magnetization reaches its first  minimum. The average frequency of the oscillations depends on the staggered field in a square-root scaling;
see the blue circle markers and their corresponding fit functions in \cref{fig:freq_stag_gap}. 
Indeed, the square-root scaling is the mathematical expression of exchange enhancement \cite{khudoy25,gomon10}. 
However, the relationship between the frequency and the gap, i.e., both the initial ($\Delta$) and the long-time value of the gap 
($\Delta_\infty$) remains linear (inset in \cref{fig:freq_stag_gap}). This was to be expected because the 
scaling between the staggered threshold field and the spin gap is quadratic $h_s^\text{thr} \propto \Delta^2$ \cite{khudoy25}.
Notably, the linear fits ${\bar{\omega}_{\mathrm{fit,3}}=p_3\cdot\Delta}$ and ${\bar{\omega}_{\mathrm{fit,4}}=p_4\cdot\Delta_\infty}$ 
show prefactors similar to those in the uniform field case.

\end{document}